\documentclass{aa}  

\usepackage{graphicx}
\usepackage{txfonts}
\usepackage{threeparttable}
   \usepackage[	colorlinks=true, 
   				citecolor=blue,
   				urlcolor=blue,
   				linkcolor=blue,
               pdftex,   % for pdflatex
               hyperindex=true,   % Make the text of index entries into hyperlinks
               bookmarks=true,   % Create bookmarks for the contents
               bookmarksopen=true,   % Expand the bookmark tree
               bookmarksnumbered=true]   % Include section numbers
               {hyperref}   % To turn references, figures, titles... into hyperlinks. 'hyperref' has to be the LAST package, except if one uses the package 'glossaries'

\begin{document}

   \title{Tomography of cool giant and supergiant star atmospheres}
   
   \subtitle{I. Validation of the method}

   \author{K. Kravchenko
          \inst{1}, S. Van Eck \inst{1}, A. Chiavassa \inst{2}, A. Jorissen\inst{1}, 
          B. Freytag \inst{3}
          \and
          B. Plez \inst{4}
          }

   \institute{Institut d'Astronomie et d'Astrophysique, Universit\'e Libre de Bruxelles,
              CP. 226, Boulevard du Triomphe, 1050 Bruxelles, Belgium\\
              \email{kateryna.kravchenko@ulb.ac.be} 
         \and
             Universit\'e C\^ote d'Azur, Observatoire de la C\^ote d'Azur, CNRS, Lagrange, CS 34229, 06304 Nice Cedex 4, France 
                 \and
             Department of Physics and Astronomy at Uppsala University, Regementsv{\"a}gen 1, Box 516, SE-75120 Uppsala, Sweden      
          \and
             Laboratoire Univers et Particules de Montpellier, Université Montpellier II, CNRS, 34095, Montpellier Cedex 05, France   
            }

   \date{Received; accepted}

  \abstract
   {Cool giant and supergiant star atmospheres are characterized by complex velocity fields originating from convection and pulsation processes which are not fully understood yet. The velocity fields impact the formation of spectral lines, which thus contain information on the dynamics of stellar atmospheres.}
   {The tomographic method allows to recover the distribution of the component of the velocity field projected on the line of sight at different optical depths in the stellar atmosphere. The computation of the contribution function to the line depression aims at correctly identifying the depth of formation of spectral lines in order to construct numerical masks probing spectral lines forming at different optical depths.}
   {The tomographic method is applied to one-dimensional (1D) model atmospheres and to a realistic three-dimensional (3D) radiative hydrodynamics simulation performed with CO5BOLD in order to compare their spectral line formation depths and velocity fields.}
   {In 1D model atmospheres, each spectral line forms in a restricted range of optical depths. On the other hand, in 3D simulations, the line formation depths are spread in the atmosphere mainly because of temperature and density inhomogeneities. Comparison of 
   %\LEt{Please spell out all acronyms the first time they appear in the paper, followed by the abbreviation in parentheses, both in the abstract and again in the main text. After that, please only use the abbreviation. See A and A language guide Section 5.2.4 www.aanda.org/language-editing}
   {cross-correlation function (CCF)} profiles obtained from 3D synthetic spectra with velocities from the 3D simulation shows that the tomographic method correctly recovers the distribution of the velocity component projected on the line of sight in the atmosphere.}
   {}

\titlerunning{Tomography of cool giant and supergiant star atmospheres}   
\authorrunning{Kravchenko et al.}  
\maketitle

%
%-------------------------------------------------------------------

\section{Introduction}
\label{Sect: Intro}

Cool giant and supergiant stars are evolved stars that play an important role in the chemical enrichment of the Galaxy. They have atmospheres with complex velocity fields, due to processes such as convection, pulsations, formation of molecules and dust, and the development of mass loss. These velocity fields impact the formation of spectral lines, the profiles of which become asymmetric, sometimes appearing double-peaked. 

\citet{Schwarzschild} suggested that the doubling of absorption lines originally observed in Cepheid variables is related to the passage of a shock wave through the photosphere. According to his scenario, in pulsating atmospheres (like those of long-period variable stars - LPVs - or Cepheids), when the shock front is located below the line-forming region, all lines are red-shifted since the matter in the line-forming region is falling down. When the shock front propagates through the line-forming region, the lines exhibit an additional blue-shifted component corresponding to rising matter. The intensity of the blue component increases with respect to the red component as the shock front reaches the upper layers of the stellar atmosphere. Finally, when the shock wave has passed through the entire photosphere, all lines exhibit only a blue-shifted component. This outward motion of the shock front was observed in Mira variables thanks to the tomographic method developed by \citet{2000A&A...362..655A,2001A&A...379..288A,2001A&A...379..305A}, which was later applied to red supergiant star atmospheres by \citet{2007A&A...469..671J} \citep[see][for a recent review]{2016ASSL..439..137J}. The method allows one to recover the distribution of the component of the velocity field projected on the line of sight at different optical depths in the stellar atmosphere. 

This paper describes a recent improvement of the tomographic technique which now resorts to the contribution function to access the depth of formation of the spectral line. 
Section~\ref{Sect: tomo-method} explains in detail the improved tomographic method and its comparison with the more basic approach proposed by \citet{2001A&A...379..288A}. Section~\ref{Sect: 3D} deals with the comparison of line formation depths in 1D and 3D model atmospheres and demonstrates the reliability of the stellar velocity field reconstruction.  

%--------------------------------------------------------------------

\section{The tomographic method}
\label{Sect: tomo-method}

%-------------------------------------------------------------------- 

The tomographic method requires three steps:
\begin{itemize}
\item computation of 1D synthetic spectra of late-type giant or supergiant stars, and, from those, identification of the depth of formation of any given spectral line;

\item construction of numerical masks selecting lines forming in selected ranges of optical depths;

\item cross-correlation of these masks with series of spectra in order to extract information on the average shape of lines forming at a given depth in the atmosphere and thus on the corresponding velocity field (i.e., the $V_z$ component of the velocity vector, where $z$ is the axis in the cartesian frame of the star pointing towards the observer).

\end{itemize}

Each of these steps is now described in turn.

%--------------------------------------------------- One column table
\begin{table*}
\begin{center}
\caption[]{Parameters of 1D MARCS model atmospheres and of the 3D simulation used in the present work.}
\label{table:1} 
\begin{threeparttable}
\begin{tabular}{c c c c c c c c}
\hline \hline
\noalign{\smallskip}
  & Model & Grid  & $T_{\rm eff}$ & $\rm \log \,g$  & Mass  & Radius &\\
  &       & [grid points] & [K] & [$\rm cm \, s^{-2}$] & [$M_\odot$] & [$R_\odot$] &\\
\noalign{\smallskip}
\hline
\noalign{\smallskip}
  & 1D-RSG       & $66$    & $3400$ & $-0.4$ & $5.0$ & $588.10$ &\\
  & 1D-AGB       & $66$    & $3500$ & $ 0.9$  & $1.5$ & $ 72.12$ &\\
  & st35gm04n38 (gray) & $401^3$ & $3414 \pm 17$\tnote{a} & $-0.39 \pm 0.01$\tnote{a} & $5.0$ & $582 \pm 5$\tnote{a} &\\

\noalign{\smallskip}
\hline
\end{tabular}
\begin{tablenotes}
\item [a] The effective temperature $T_{\rm eff}$, surface gravity $\rm \log \,g$ and stellar radius of the 3D model represent averages over both spherical shells and time; errors are one-standard-deviation fluctuations with respect to the average over time \citep[see][]{2009A&A...506.1351C,2011A&A...535A..22C}. 
\end{tablenotes}
\end{threeparttable}
\end{center}
\end{table*}

\subsection{The 1D contribution function}
\label{Sect: 1DCF}

The original implementation of the method \citep{2000A&A...362..655A,2001A&A...379..288A,2001A&A...379..305A} was based on the Eddington-Barbier approximation to define the depth of line formation, stating that the emergent flux at wavelength $\lambda$ is supposed to come from the layer located at the optical depth $\tau_{\lambda}=2/3$. Thus, the line depression was assumed to form at that optical depth. However, \citet{1986A&A...163..135M} showed that the Eddington-Barbier approximation gives a valid depth of line formation only for sufficiently strong lines. In order to correctly assess the depth of line formation, {the contribution function to the absolute line depression (CFLD)} computed on a single ray (intensity) or disk-integrated (flux) has to be computed instead.
This CFLD is different from the contribution function to the relative line depression derived by
\citet{1986A&A...163..135M}.

Following \citet{1986A&A...163..135M}, \citet{1996MNRAS.278..337A} derived the ${\rm CFLD_{SR}}$ 
(for a single ray) in the specific intensity, with $Q\equiv I_c-I_l$:% (here for the direction towards the observer): 

\begin{equation}
{\rm CFLD_{SR}}(\log \tau_0,{\lambda}) = (\ln10) \, \mu^{-1} \frac{\tau_0}{\kappa_{c,0}} \kappa_{l_{\lambda}} \,(I_{c_{\lambda}} - S_{l_{\lambda}}) \, e^{-\tau_{\lambda}/\mu}  
\label{Eq:cfmagain}
,\end{equation}
\ 
and in the flux by integrating on the disk (DI), considering a plane parallel atmosphere:
\
\begin{equation}
{\rm CFLD_{DI}}(\log \tau_0,{\lambda}) = {2\pi}\,(\ln10) \frac{\tau_0}{\kappa_{c,0}} \int_0^1 \kappa_{l_{\lambda}} \,(I_{c_{\lambda}} - S_{l_{\lambda}}) \, e^{-\tau_{\lambda}/{\mu}} \, d\mu 
\label{Eq:cfalbrow}
,\end{equation}
\
with
\[
\begin{array}{lp{0.7\linewidth}}
\tau_0 = \int \kappa_{c,0} \, \rho \, {\rm d}x           & the continuum optical depth at a reference wavelength $\lambda_0$ ($\rho$ is the density). The $\tau_0$ scale can be adopted as a proxy to the geometrical depth $x$;\\
%The $\tau_0$ scale can be adopted because it is proportional to the geometrical depth $x$ (at least when $\kappa_{c,0}$ and $\rho$ may be considered %constant along the ray);\\
\kappa_{c,0}         & the continuum absorption coefficient at a reference wavelength $\lambda_0$;\\
I_c              & the continuous intensity;\\
I_l                              & the line intensity; \\
\kappa_l         & the line absorption coefficient;\\
S_l              & the line source function;\\
\tau_{\lambda} = \int \kappa_{\lambda} \, \rho \, {\rm d}x            & the optical depth along the ray ($\kappa_{\lambda}$ is the total absorption coefficient at wavelength $\lambda$);\\
\mu=\cos{\theta} & the cosine of the angle $\theta$ between the line of sight and the radial direction ($\mu = 1$ for the direction towards the observer).\\
\end{array}
\]
\
The optical depth $\tau_{\lambda}$ is related to the reference optical depth $\tau_0$ through

\begin{equation}
\frac{{\rm d}\tau_{\lambda}}{{\rm d}\tau_0} = \frac{\kappa_{\lambda}}{\kappa_{c,0}}.
\label{Eq:tau}
\end{equation}

%                                                One column figure
%----------------------------------------------------------------- 
\begin{figure}[h!]
\centering
\includegraphics[width=9cm]{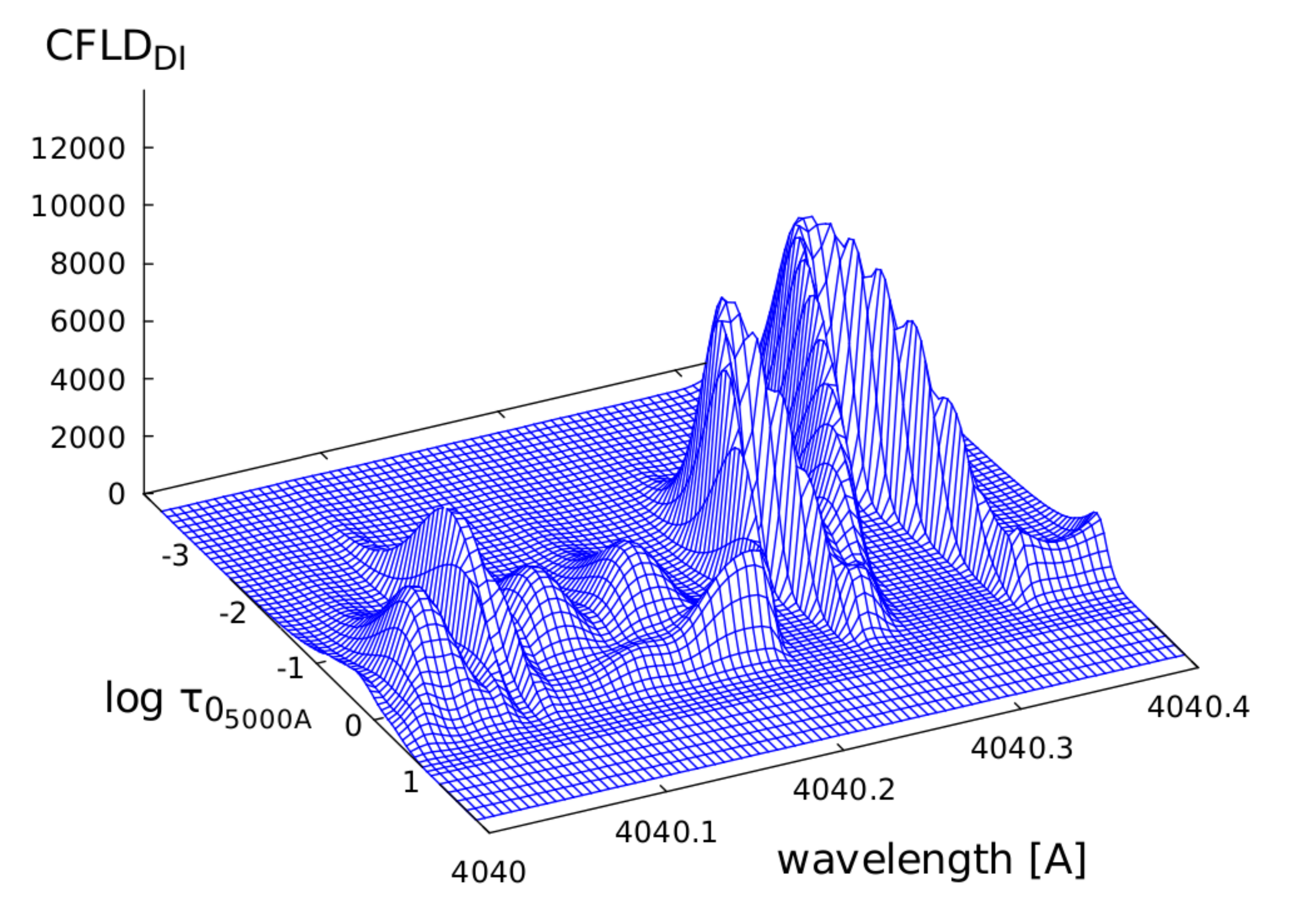} \\     
\includegraphics[width=8.5cm]{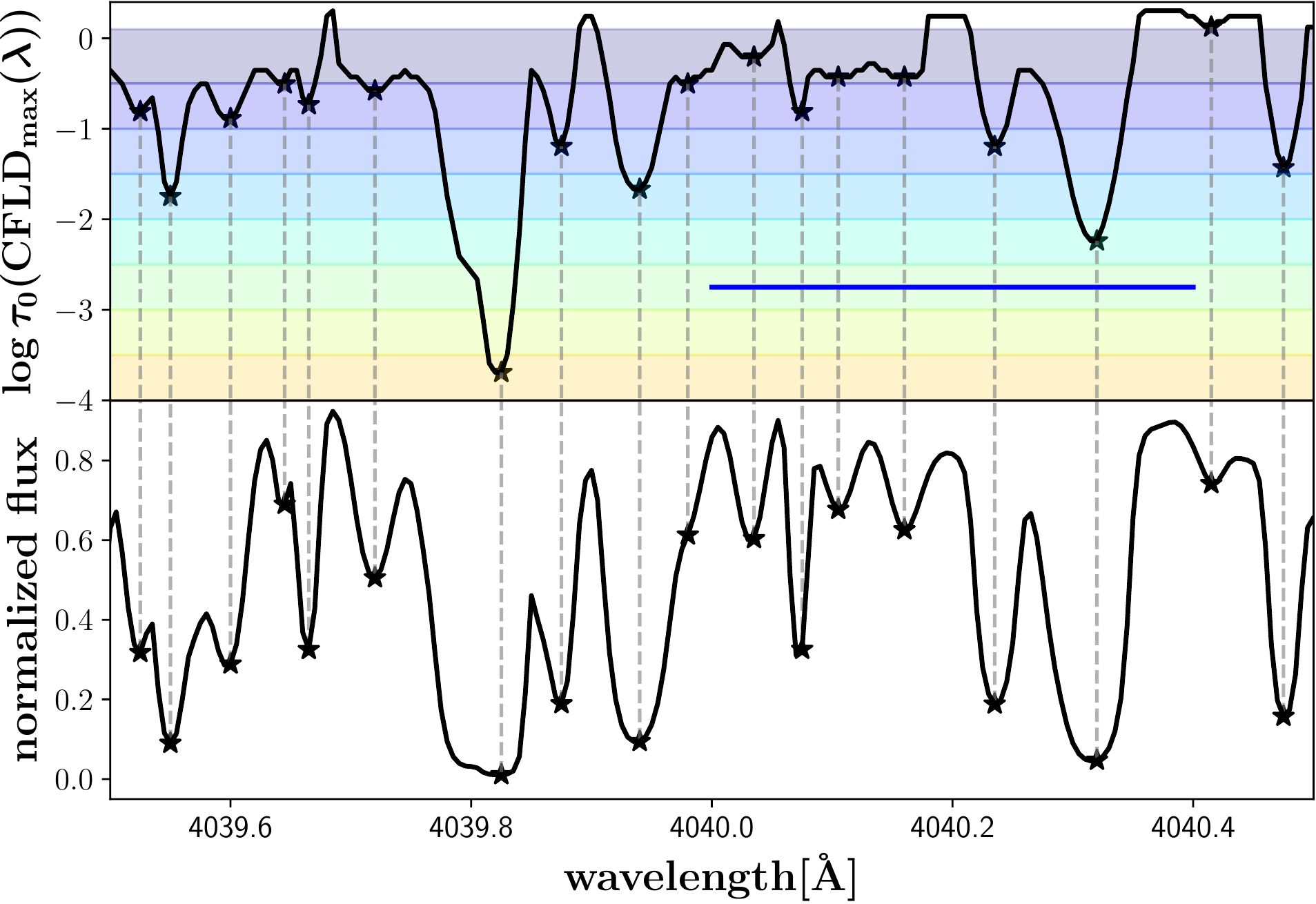}
\caption{\textit{Top panel:} The $\rm CFLD_{DI}$ for the 1D-RSG model (Table~\ref{table:1}). \textit{Middle panel:} The depth function $D(\lambda) \equiv \log \tau_0 (\rm CFLD_{max} (\lambda))$ corresponding to the crest line of the CFLD in the $\lambda - \log \tau_0$ plane. Star symbols correspond to minima of the depth function. The horizontal bands represent the $\log \tau_0$ ranges spanned by the different masks. The horizontal blue line identifies the spectral interval of the displayed CFLD (top panel). \textit{Bottom panel:} The corresponding synthetic spectrum.}
\label{fig:1Dcfsurface}
\end{figure}
%-----------------------------------------------------------------

\subsection{Mask construction}
\label{Sect: mask-construction}

All the necessary quantities for the CFLD computation have been extracted from the radiative-transfer code TURBOSPECTRUM \citep{2012ascl.soft05004P}. The geometry of the 1D radiative transfer is explained in \citet{2008A&A...486..951G}. 
The TURBOSPECTRUM code was previously modified by \citet{2013EAS....60...85L} in order to include the computation of the CFLD. The radiative transfer was deliberately performed with zero microturbulence velocity (see Sect.~\ref{Sect: 3D-single-ray}).

The top panel of Fig.~\ref{fig:1Dcfsurface} displays the $\rm CFLD_{DI}$ (computed with Eq.~\ref{Eq:cfalbrow}) for a 1D MARCS model atmosphere with $T_{\rm eff}=$ 3400 K and $\log g =$ -0.4 (1D-RSG model of Table~\ref{table:1}). Its local maximum value (for a given wavelength $\lambda$) is a function of the reference optical depth $\tau_0$. The reference optical depth scale $\tau_0$ is computed at 5000 $\AA$ using only continuum opacities. The set of $\tau_0$ corresponding to the maximum CFLD for each wavelength defines a crest line. We call this crest line the "depth function" $D(\lambda)$ (in the $\log \tau_0$ scale, middle panel of Fig.~\ref{fig:1Dcfsurface}); it specifies the optical depth $\tau_0$ at which the line(s) contributing at wavelength $\lambda$ mainly form(s).  
The comparison of the depth function with a synthetic spectrum (bottom panel of Fig.~\ref{fig:1Dcfsurface}) shows that, as expected, the cores of spectral lines form in outer layers while wings form in deeper layers. 

%                                                One column figure
%----------------------------------------------------------------- 
\begin{figure}[h]
\centering
\includegraphics[width=5.5cm]{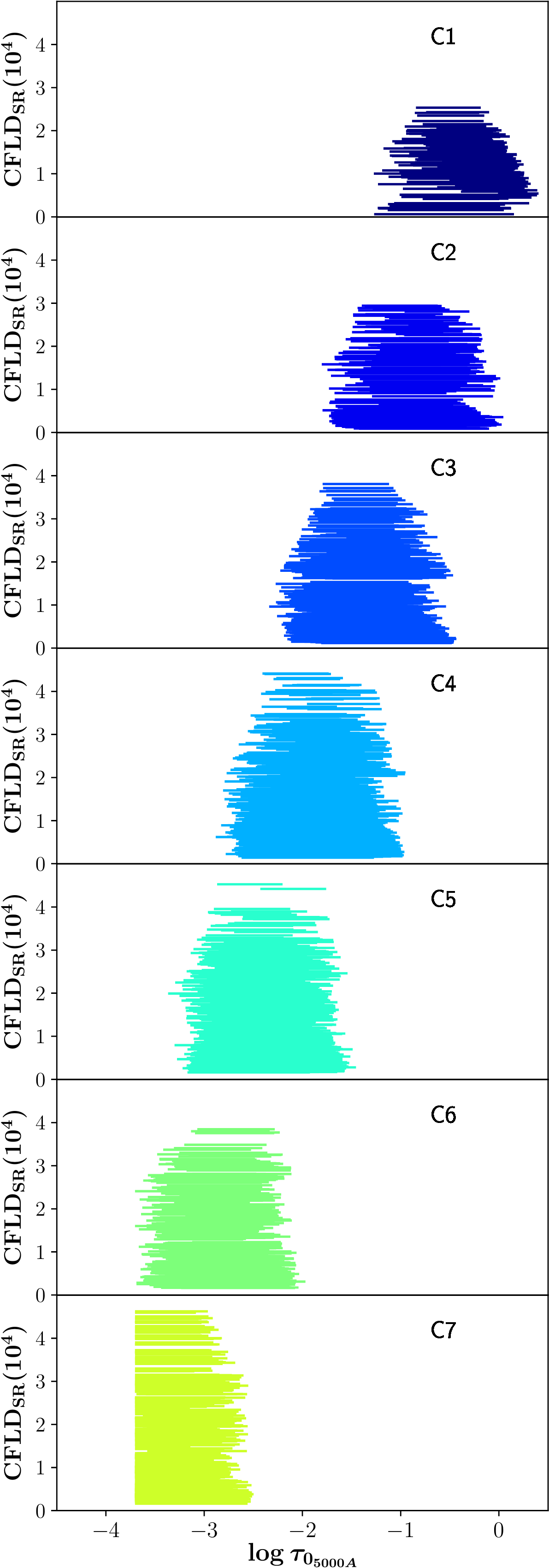}
\caption{The FWHM of the $\rm CFLD_{DI}$ for lines contributing to spectral masks C1-C7 (C8 not represented, see text) computed from the 1D-RSG model of Table~\ref{table:1}. The color-code identifies the masks as in the middle panel of Fig.~\ref{fig:1Dcfsurface}.}
\label{fig:fwhm}
\end{figure}
%----

Once the depth function is computed, the atmosphere is split into different slices (horizontal bands in the middle panel of Fig.~\ref{fig:1Dcfsurface}) and spectral masks are constructed for each slice. Each mask is a collection of Dirac-like distributions. A set of eight tomographic masks were built. Each mask contains the positions of all lines whose depth function minimum falls within a limited range of optical depths. The optical depth full range is chosen as $-4.0 \leq \log\tau_0 \leq 0.0$ with steps 0.5 in units of logarithmic optical depth. Following \citet{2001A&A...379..288A}, only wavelengths where atomic lines dominate are kept in masks to achieve the best wavelength accuracy, which cannot always be reached with molecular lines.

%                                                One column figure
%----------------------------------------------------------------- 
\begin{figure}
\centering
\includegraphics[width=8.5cm]{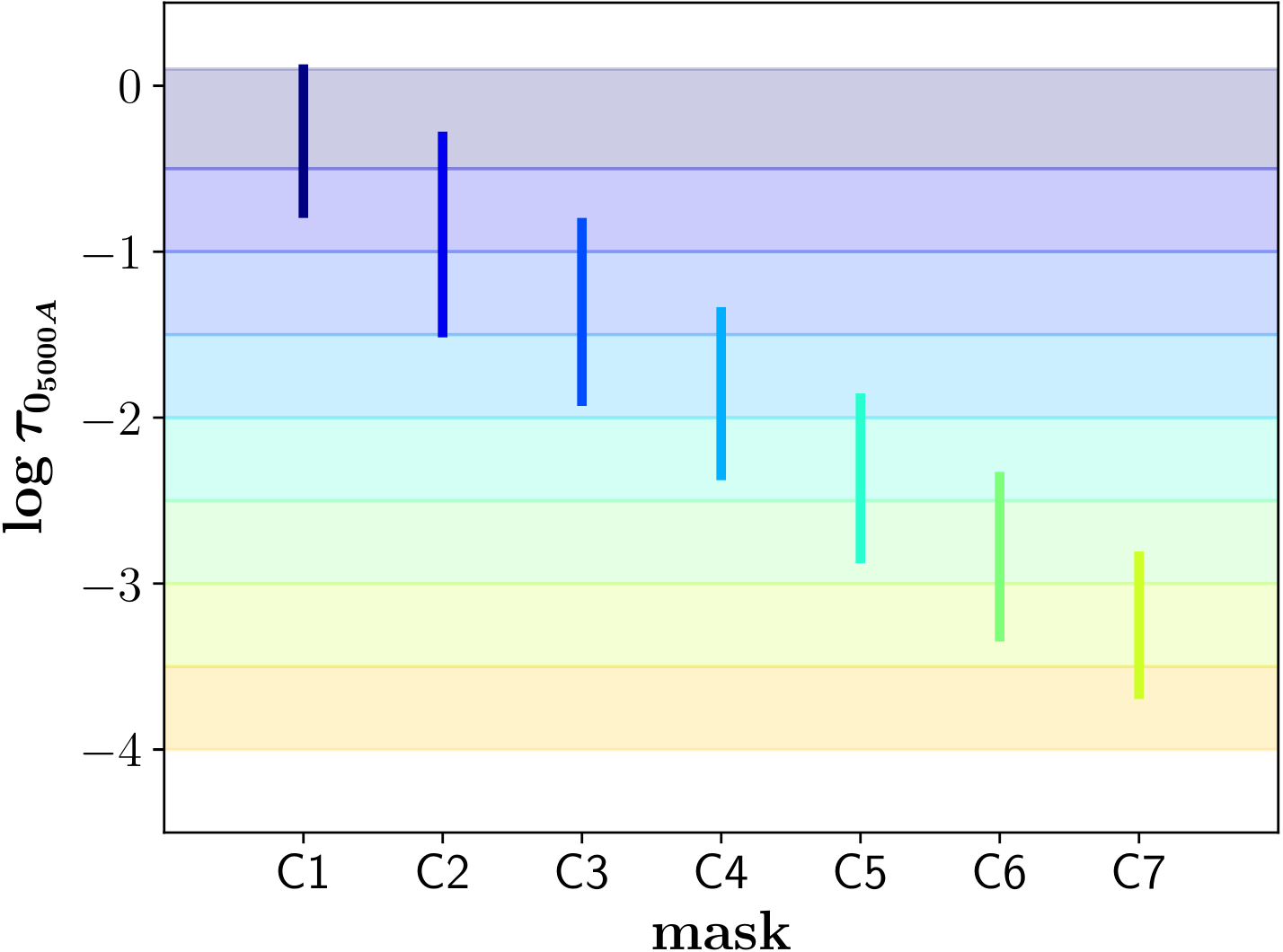}
\caption{\textit{Vertical lines:} Average FWHM of the CF of lines contributing to spectral masks C1-C7 (C8 not represented, see text) computed from the 1D-RSG model of Table~\ref{table:1}. \textit{Horizontal bands:} Optical depth ranges used for selecting lines in a given mask. 
The color code has no spectral meaning. }
\label{fig:fwhmaverage}
\end{figure}
%-----------------------------------------------------------------

Masks constructed in this way are then cross-correlated with either observed or synthetic stellar spectra. The resulting {CCF} provides information about projected velocities (position of the minimum of the CCF), average strength (depth of the CCF) and shape of lines in a given atmospheric layer.

To test how well a given mask from the series of Fig.~\ref{fig:1Dcfsurface} probes lines formed in a given layer, the full width at half maximum (FWHM) of CFLD profiles (along the $\tau_0$ axis) of all lines contributing to the masks were computed. They are displayed in Fig.~\ref{fig:fwhm}. The outermost mask C8 is not shown because its CFLD profiles are truncated due to the lack of extension of the 1D model atmosphere towards low optical depths (below $\log\tau_0 = -3.7$), and the FWHM could not be computed. Figure~\ref{fig:fwhmaverage} displays the average FWHM for each mask as a vertical line. It is seen that, with the chosen step of 0.5 in $\log \tau_0$, neighboring masks have overlapping FWHM of their CFLD profiles. To probe the velocity field in non-overlapping layers, one may consider the set of masks \{C1, C3, C5, C7\} or \{C2, C5, C7\} (Fig.~\ref{fig:fwhmaverage}). However, this cross-talk between masks is moderate; indeed in Sect.~\ref{Sect: V-recovery} we will demonstrate that using all masks does allow us to reliably reconstruct the projected velocity field.

\subsection{Comparison of "old" and "new" tomographic techniques: the Mira variable V Tau}

\label{Sect: Comparison}

To test the tomographic method based on the CFLD, we first tried to reproduce the results obtained by \citet{2001A&A...379..288A} for the Mira variable star V Tau. 

For this purpose, another set of spectral masks was constructed with the same hypotheses as in \citet{2001A&A...379..288A}, except for the Eddington-Barbier condition replaced by the following: the core of a given spectral line is forming in the layer where the depth function (i.e., the maximum of the CFLD, computed with Eq.~\ref{Eq:cfalbrow}, along the wavelength axis) reaches the lowest optical depth (identified by star symbols in the middle panel of Fig.~\ref{fig:1Dcfsurface}). The construction of the spectral masks proceeds as in \citet{2001A&A...379..288A}:

\begin{itemize}
\item A 1D MARCS model atmosphere is used with $T_{\rm eff}$ = 3500 K and $\log g$ = 0.9 (1D-AGB model of Table~\ref{table:1}). It must be mentioned that the model atmosphere used by \citet{2001A&A...379..288A} was computed with the \citet{1992A&A...256..551P} version of the MARCS code whereas the model used here is from the more recent MARCS grid \citep{2008A&A...486..951G}; 
\item Linelists are used containing data for the same atoms and molecules, and identical CNO abundances and isotopic ratios;
\item A reference optical depth scale at 1.2 $\mu \rm m$ is adopted; 
\item The atmosphere is divided into eight layers in the range $-8.00 \leq \log\tau_{0_{1.2\mu \rm m}} \leq -2.00$ with step 0.75 in units of logarithmic optical depth; 
\item The masks-only wavelengths corresponding to atomic lines are kept.

\end{itemize}

\begin{figure}
\centering
\includegraphics[width=8cm]{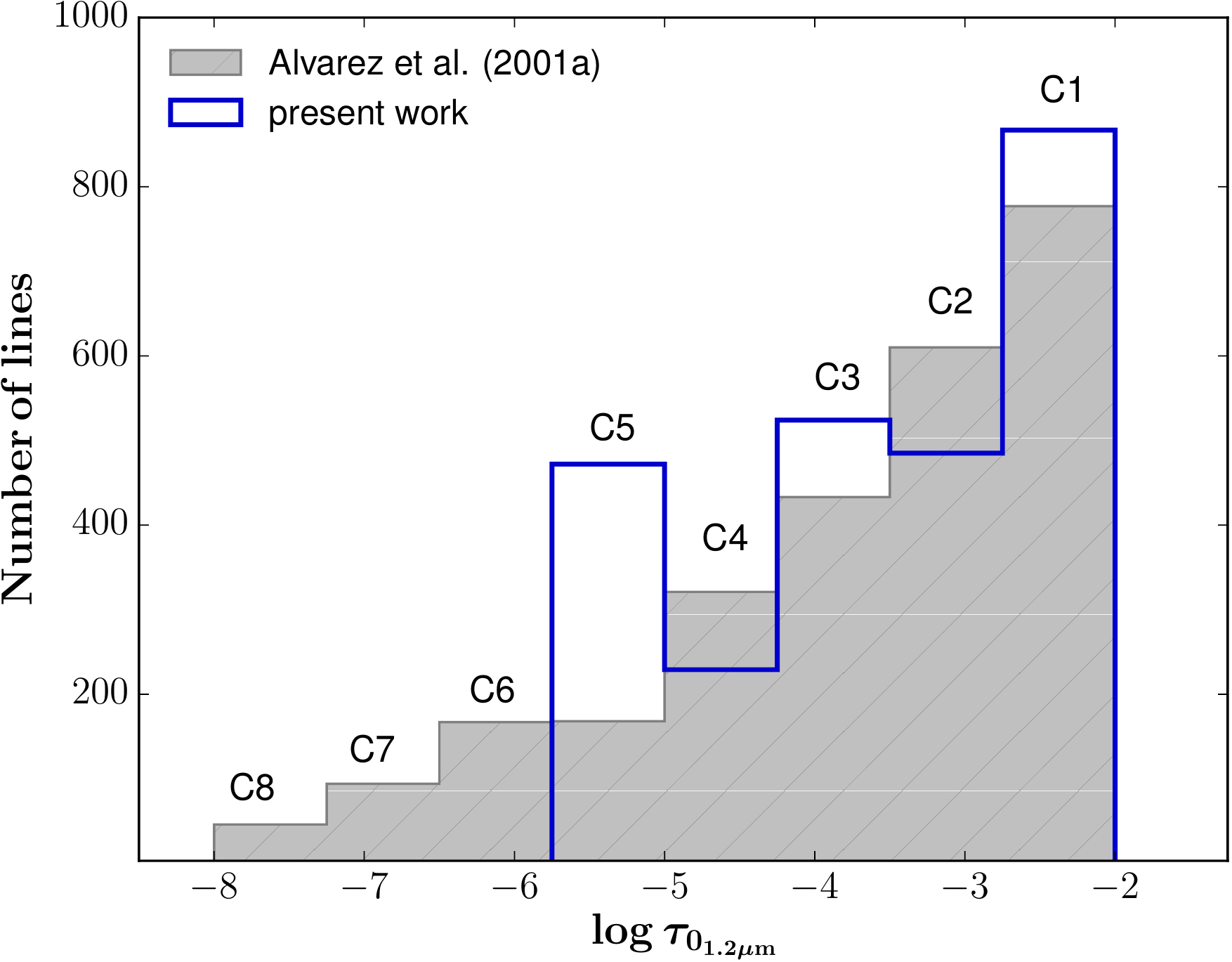}
\caption{The distribution of lines in spectral masks C1-C8 obtained by \citet{2001A&A...379..288A} (gray shaded histogram) and in the present work (blue line) for the 1D-AGB model (Table~\ref{table:1}).}
\label{fig:4}
\end{figure}

Figure~\ref{fig:4} shows the number of lines in the various masks together with the distribution of lines in masks from \citet{2001A&A...379..288A}; it illustrates the fact that our MARCS model atmosphere does not extend as far as the model used by \citet{2001A&A...379..288A}. Therefore, all lines which form in the optical depth range $-8 < \log\tau_{0_{1.2 \mu \rm m}}<-5.75$ (masks C6-C8) in \citet{2001A&A...379..288A} are merged in our mask C5.

\begin{figure*}
\centering

\includegraphics[width=18.5cm]{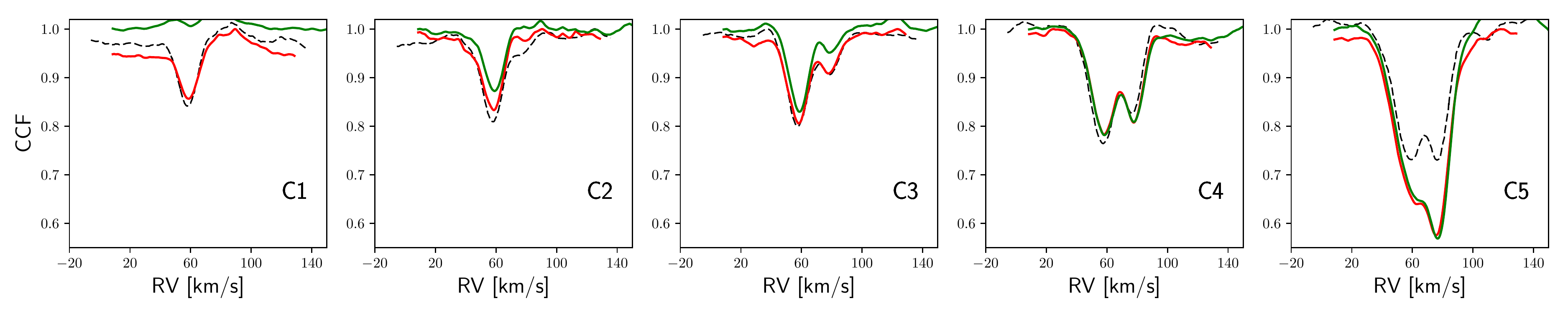}
\caption{\textit{Black dashed line:} Sequence of CCFs obtained from a V Tau spectrum (JD 2451093.5) using masks of  \citet{2001A&A...379..288A} (their Fig. 18). \textit{Red line:} the CCF profiles obtained from the same spectrum  
using masks with identical $\log \tau$ limits but built using maxima of the {CFLD}, as described in Sect.~\ref{Sect: mask-construction}. \textit{Green line:} the CCF profiles obtained from the same spectrum using masks built from Eq.~\ref{Eq:averagetau} in Sect.~\ref{Sect: meanformation-depth}.}
\label{fig:5}
\end{figure*}

The obtained masks were cross-correlated with an ELODIE \citep{1996A&AS..119..373B} spectrum of V Tau. Figure~\ref{fig:5} compares the resulting CCFs to those of \citet{2001A&A...379..288A}. The discrepancy between CCFs in the outermost mask C5 is due to different extensions of the model atmospheres described above. Our CCFs show very good agreement with those obtained by \citet{2001A&A...379..288A}. The comparison with our more sophisticated method based on the CFLD indicates that the approach of \citet{2001A&A...379..288A} (using the Eddington-Barbier condition) yields meaningful results. It was indeed able to reveal the occurrence of the Schwarzschild scenario in Mira variables.

%                                                One column figure
%----------------------------------------------------------------- 
%----

%                                                One column figure
%----------------------------------------------------------------- 
%----

\subsection{Mean formation depth of spectral lines}
\label{Sect: meanformation-depth}

Instead of constructing masks from the maximum of the CFLD, one may use the average depth of line formation obtained by the first moment of the {CFLD}.
For any given wavelength, \cite{1986A&A...163..135M} defines  \citep[his Eq. 22, also used in][]{2015MNRAS.452.1612A}:

\begin{equation}
<x> = \frac{\int x~ CFLD(x) ~dx}{\int CFLD(x) ~dx},
\label{Eq:averagetau}
\end{equation}
\
with $x = \log \tau_0$, that is the logarithmic reference optical depth. 

A set of five masks was constructed as in Sect.~\ref{Sect: Comparison}, and the $\log \tau_0$ corresponding to the maximum of the CFLD was replaced by the average line formation depth computed with Eq.~\ref{Eq:averagetau}. The resulting masks were then cross-correlated with the observed spectrum of V Tau. The resulting CCFs are shown in Fig.~\ref{fig:5} (green line) together with those obtained in Sect.~\ref{Sect: Comparison} from the maxima of the {CFLD}. They are characterized by noisy and less contrasted profiles in masks C1-C3. 

Actually, the depth of line formation, as derived either from the extrema of the CFLD or from a weighted average,  will differ in the case of skewed {CFLDs}, as is mostly the case for weak lines forming deep in the atmosphere.  Because of skewed CFLD, the averaging method (Eq.~\ref{Eq:averagetau}) will attribute to weak lines a formation depth higher up in the atmosphere. Thus, they will be included in masks also containg stronger lines often characterized by more symmetric {CFLDs} and forming naturally higher up in the atmosphere. As a result, the inner masks will contain fewer lines and give rise to noisy and less contrasted CCFs.  

Thus, we adopted the approach based on the maximum of the CFLD (the so-called depth function) which also provides results  that are more consistent with \citet{2001A&A...379..288A}.

\section{Application to 3D atmospheres of red supergiants}
\label{Sect: 3D}

The tomographic method uses 1D model atmospheres for mask construction. However, in real stars like Mira and supergiant stars, multi-dimensional dynamical processes occur, and \citet{2011A&A...535A..22C} showed that more than one hydrostatic 1D model is necessary to reproduce the thermodynamical structure of the radiative hydrodynamics (RHD) simulation.  
In such complex atmospheres, the depths of formation of spectral lines will likely differ from the 1D prediction. 

In order to assess the reliability of the tomographic method in 3D dynamical atmospheres, it was tested on snapshots from numerical 3D RHD simulations performed with the CO5BOLD code \citep{2012JCoPh.231..919F}. The 3D simulations are characterized by realistic input physics and reproduce the effects of convection and non-radial waves \citep{2011A&A...535A..22C}. 
The parameters of the 3D simulation used in the analysis are presented in Table~\ref{table:1}.

%                                                One column figure
%----------------------------------------------------------------- 
\begin{figure}[h]
\centering
\includegraphics[width=7.6cm]{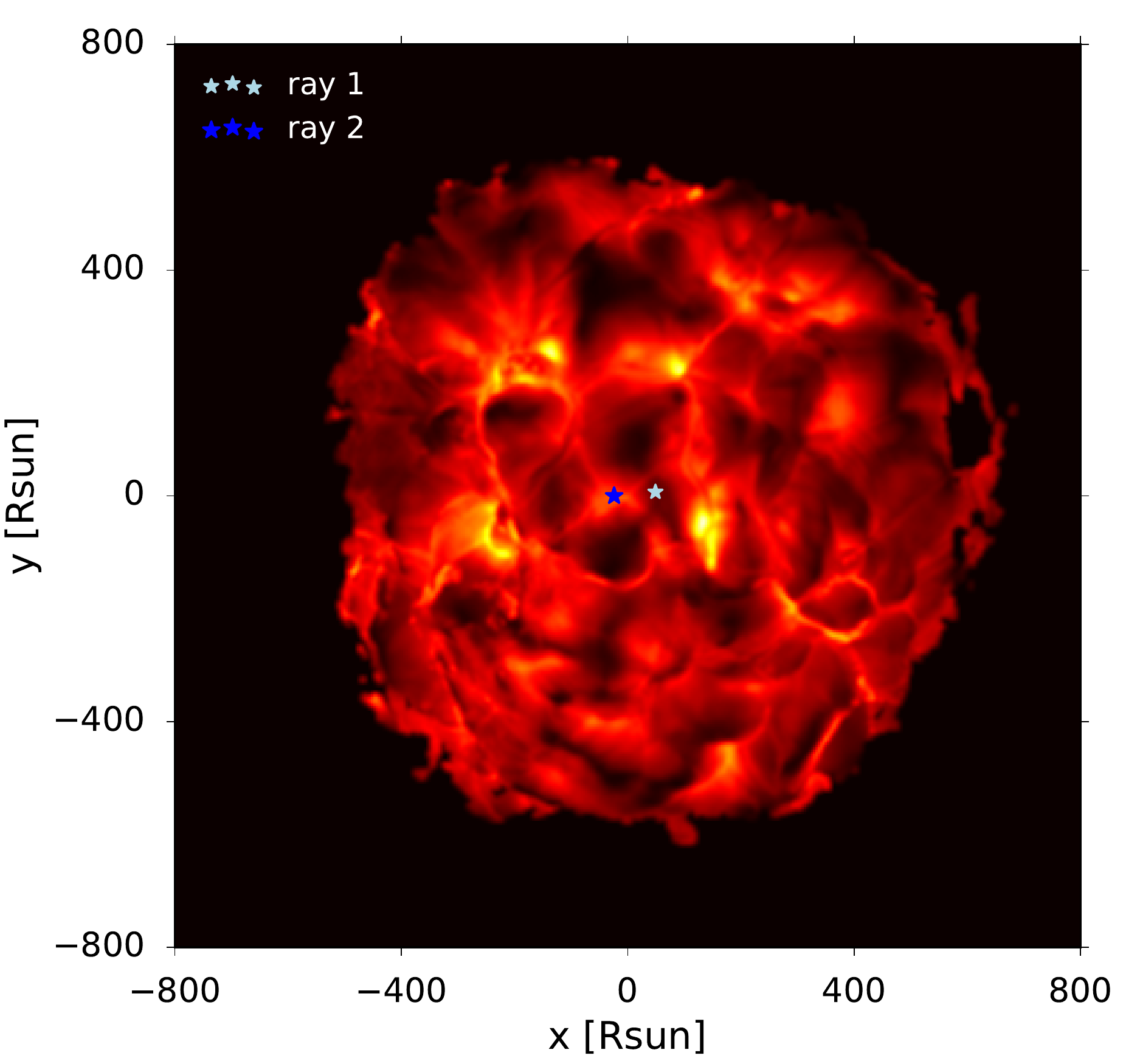}
\caption{The intensity map at $\lambda =$ 4040.07 $\AA$ for a snapshot of the 3D simulation. The intensity range is [0; 250] $\rm erg \, cm^{-2} s^{-1} \AA^{-1}$. Rays 1 and 2 are analyzed in Sect.~\ref{Sect: 3D-single-ray}.}
\label{fig:6}
\end{figure}
%----

The 3D pure-LTE radiative transfer code Optim3D \citep{2009A&A...506.1351C} was used to compute synthetic spectra and intensity maps from a snapshot of the 3D simulations (Fig.~\ref{fig:6}). The code takes into account the Doppler shifts caused by the convective motions. The radiative transfer is calculated using pre-tabulated extinction coefficients generated with the MARCS code \citep{2008A&A...486..951G}. These tables are functions of temperature, density, and wavelength, and are computed by adopting the solar composition of \citet{2006NuPhA.777....1A}. They include the same extensive atomic and molecular data as the MARCS models used in the present work. 

The computation of the {CFLD} was implemented in the OPTIM3D code. The 3D $\rm CFLD_{SR}$ and 3D $\rm CFLD_{DI}$ are described in Sects.~\ref{Sect: 3D-single-ray} and ~\ref{Sect: 3D-stellar-disk}, respectively.

%                                                One column figure
%----------------------------------------------------------------- 
\begin{figure}%[h]
\centering
\includegraphics[width=8.5cm]{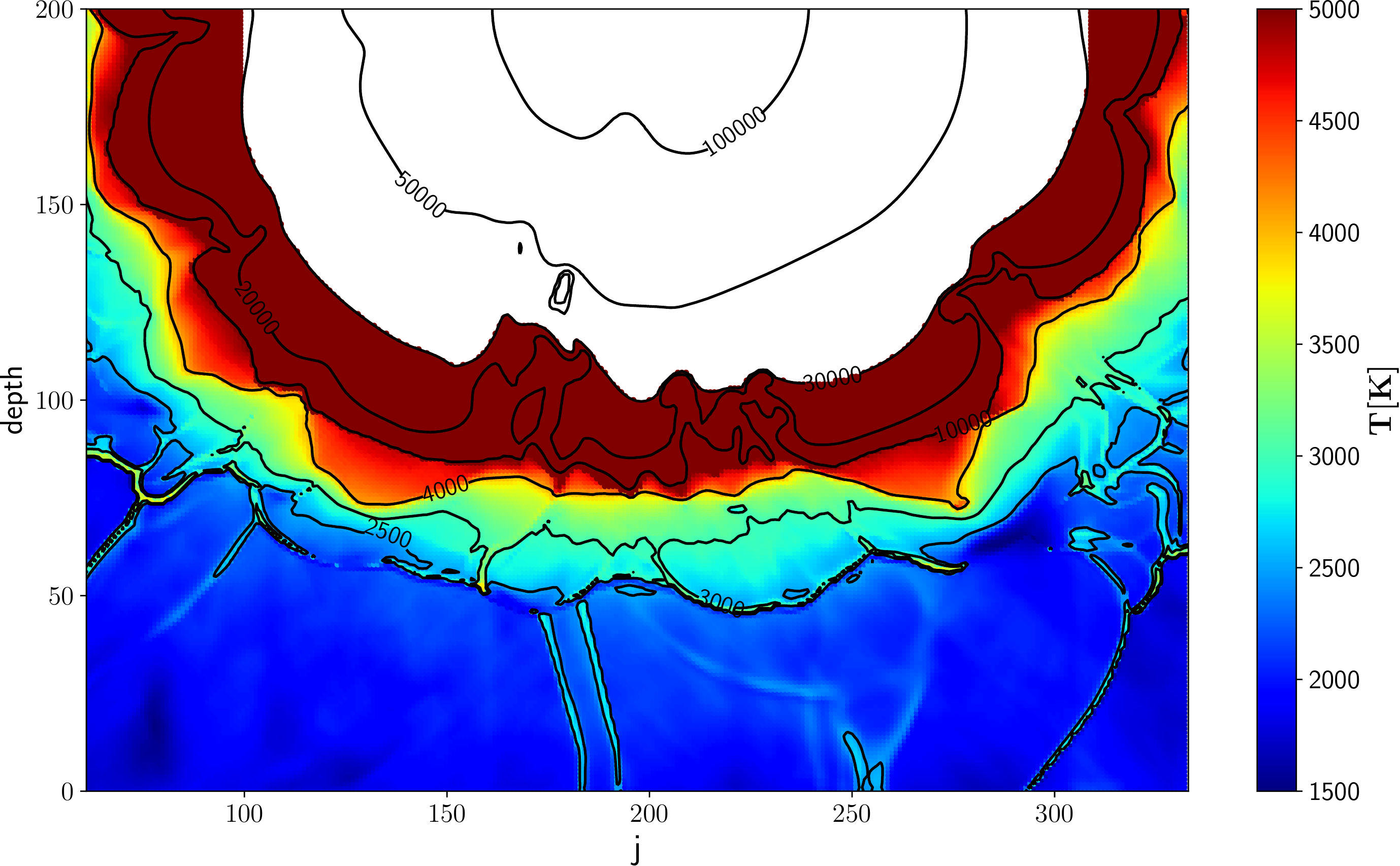} \\
\includegraphics[width=8.5cm]{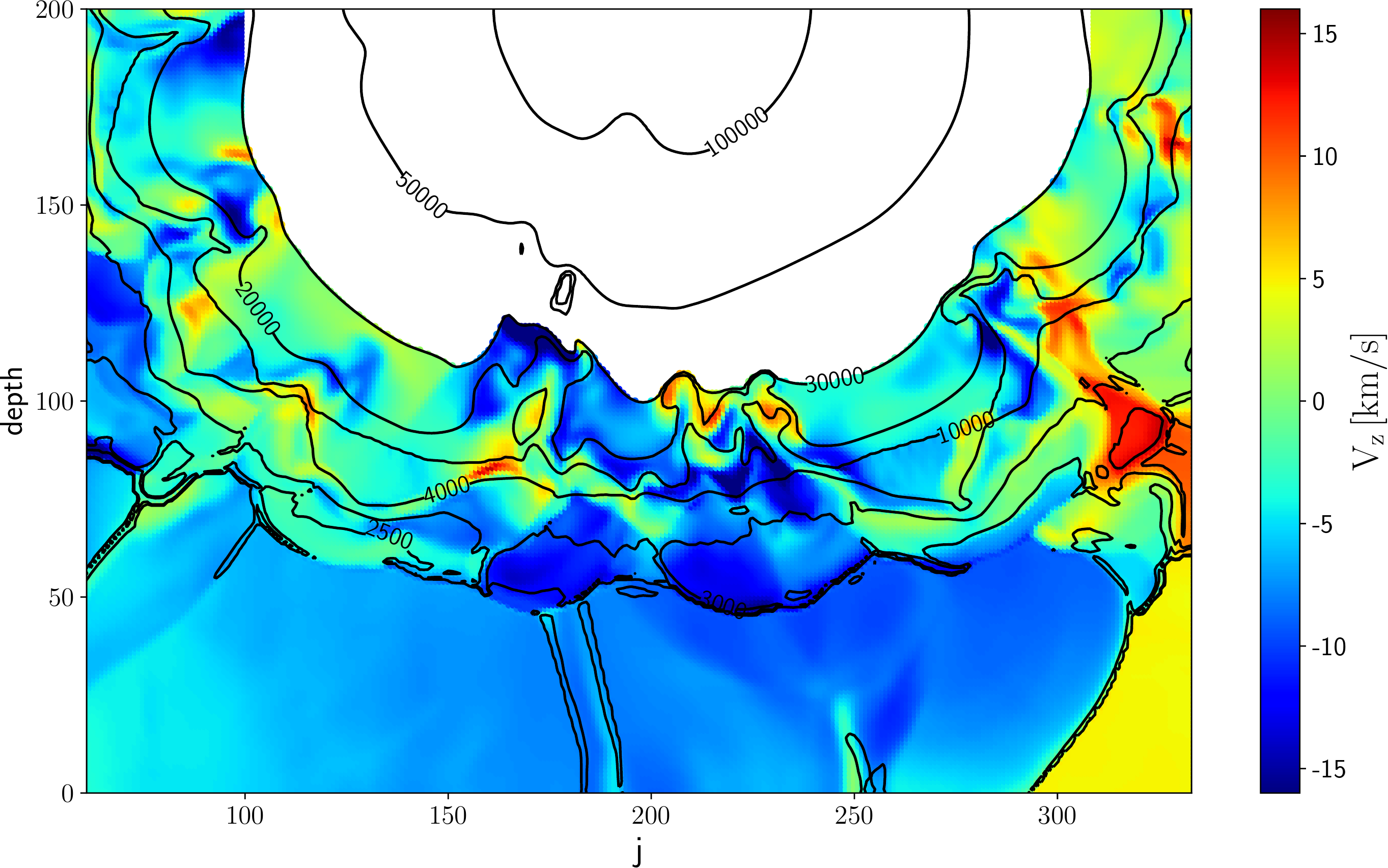} \\
\includegraphics[width=8.5cm]{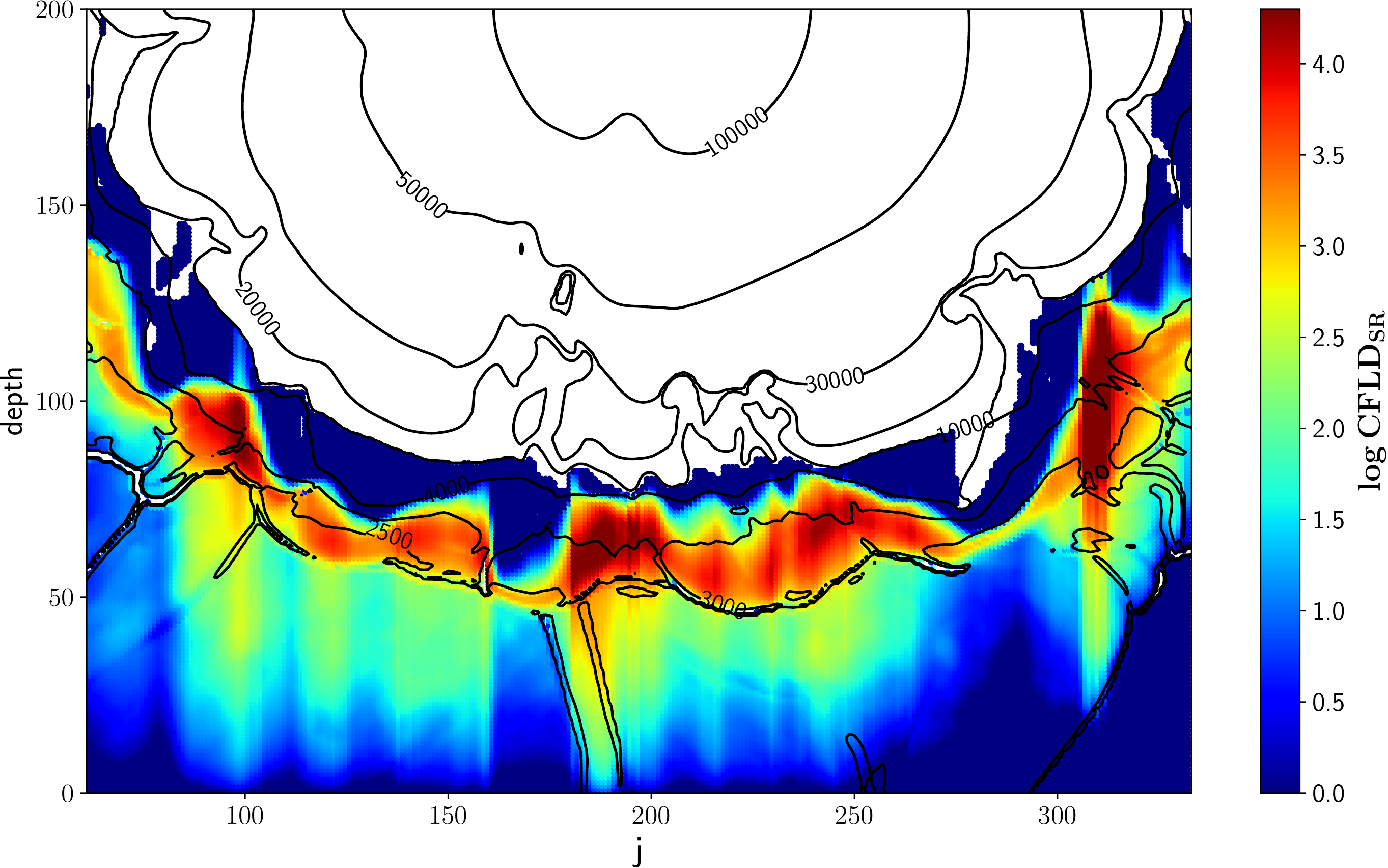} 
\caption{The temperature $T$, projected velocity $V_z$ and $\rm \log CFLD_{{SR}}$ structures for the slice through the 3D numerical box. The observer is at the bottom of each Figure. Black contour lines on all panels show the thermal stratification.}
\label{fig:7}
\end{figure}
%----

\subsection{The CFLD for the 3D simulation: single ray}
\label{Sect: 3D-single-ray}

If we consider a single ray of the 3D simulation, the 3D $\rm CFLD_{SR}$ is computed using Eq.~\ref{Eq:cfmagain} for the direction towards the observer, that is with $\mu = 1$. 

Figure~\ref{fig:7} displays the temperature, $V_z$ , and 3D $\rm CFLD_{SR}$ structures for a representative slice through the 3D numerical box. The 3D $\rm CFLD_{SR}$ was computed at $\lambda =$ 4040.07 $\AA$ (corresponding to an Ir I atomic line). The Ir I line was selected arbitrarily and contributes to the C1 mask computed in Sect.~\ref{Sect: mask-construction}. Figure~\ref{fig:7} demonstrates the complexity of a 3D atmosphere with respect to a 1D atmosphere. Here, the velocity field is not homogeneous but is characterized by multiple shocks which affect the temperature structure and the 3D CFLD.  
Thus, the CFLD in the 3D atmosphere is not smooth like in 1D.

\begin{figure}
\centering
\includegraphics[width=9cm]{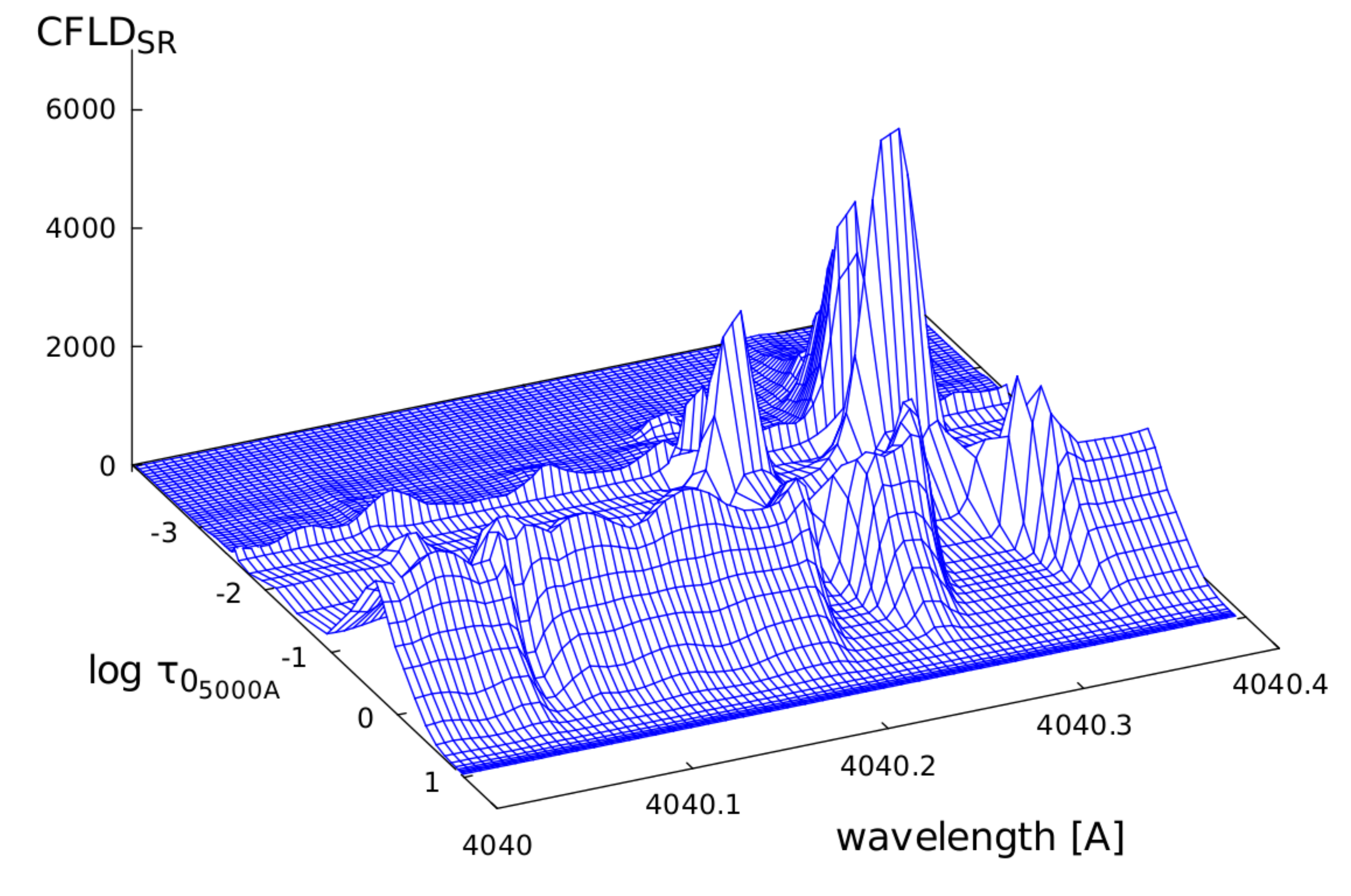}
\caption{The 3D $\rm CFLD_{SR}$ for a single ray of the 3D simulation (ray 1 on Fig.~\ref{fig:6}) in the same spectral range as the 1D $\rm CFLD_{DI}$ in the top panel of Fig.~\ref{fig:1Dcfsurface}. A spectral resolution of $R = 900000$ was adopted for plotting purposes.}
\label{fig:8}
\end{figure}

A more detailed study of the 3D $\rm CFLD_{SR}$ is performed for a single ray of the 3D numerical box (ray 1 in Fig.~\ref{fig:6}). The velocities of the 3D simulation were set to zero in the Optim3D code to have a direct comparison with hydrostatic 1D calculations (also in Sects.~\ref{Sect: 3D-stellar-disk} and ~\ref{Sect: tomo-masks}). The extinction coefficients used in Optim3D were constructed with no micro-turbulent broadening. The same approach was applied in the 1D radiative transfer computation where the microturbulence was set to zero in order to compare the CFLDs without any fudge parameter which mimics the velocity field at small scales. Thus, the velocity information will emerge only from the 3D simulation itself (see Sect.~\ref{Sect: V-recovery}). The continuum optical depth scale at 5000 $\AA$ was taken as a reference scale. The 3D $\rm CFLD_{SR}$ is shown in Fig.~\ref{fig:8}. It has a more complicated structure than the 1D $\rm CFLD_{DI}$ (top panel of Fig.~\ref{fig:1Dcfsurface}).
The 3D $\rm CFLD_{SR}$ shows multiple maxima at a given wavelength, which means that a given spectral line forms at several depths in the atmosphere.

%                                                One column figure
%----------------------------------------------------------------- 
%----

\begin{figure}%[h]
\centering
\includegraphics[width=8.4cm]{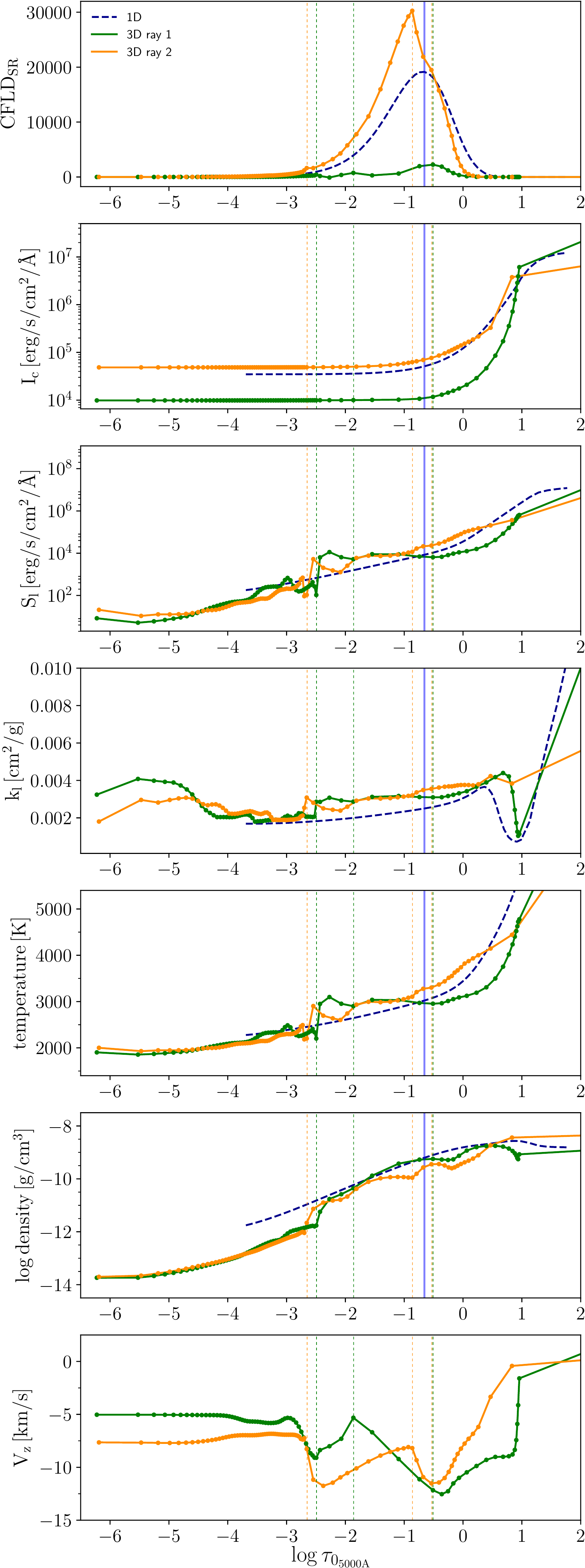}
\caption{The $\rm CFLD_{SR}$, continuum intensity $I_c$, line source function $S_l$, line absorption coefficient $k_l$, temperature, logarithmic density and $V_z$ structures at $\lambda = \rm{4040.07}~\AA$ for two single rays of the 3D snapshot located in bright and dark areas on the stellar disk (orange and green colors indicate rays 2 and 1, respectively, in Fig.~\ref{fig:6}) and for the corresponding 1D-RSG model of Table~\ref{table:1} (blue dashed line). Vertical lines correspond to local maxima of CFLDs.}
\label{fig:9}
\end{figure}

The top panel of Fig.~\ref{fig:9} displays the 3D $\rm CFLD_{SR}$ as a function of $\tau_0$ at $\lambda =$ 4040.07 $\AA$ for two single rays of the 3D simulation and for the corresponding 1D MARCS model atmosphere (i.e., 1D-RSG model of Table~\ref{table:1}).  
The rays were selected close to the center of the stellar disk and correspond to dark and bright areas on the surface (rays 1 and 2 in Fig.~\ref{fig:6}, respectively). The location of rays has an impact on the height of the 3D CFLD profile.  
The temperatures probed by the ray in a dark region (ray 1; green line in Fig.~\ref{fig:9}) are generally cooler than those probed by the ray located in a bright region (ray 2; orange line in Fig.~\ref{fig:9}). 
Thus, its continuum intensity (used in Eq.~\ref{Eq:cfmagain} and displayed in the second panel of Fig.~\ref{fig:9}) and the CFLD are also lower. Figure~\ref{fig:9} also compares the temperature (fifth panel), density (sixth panel) and $V_z$ velocity (seventh panel) for single rays of the 3D simulation and for the corresponding 1D model atmosphere. 

The 1D model is characterized by a monotonic and continuous growth of the temperature and density towards the stellar interior. On the contrary, the corresponding 3D profiles show non-monotonous temperature and density variations, including shocks, and complex velocity variations.  
The presence of a shock is indicated by a strong negative gradient in the velocity ${\mathrm d} V_z / {\mathrm d} \log \tau_0$ (see the bottom panel of Fig.~\ref{fig:9}), that signals compression of outward moving ($V_z<0$) gas.
The corresponding atmospheric layers are characterized by temperature and density discontinuities. 

The temperature and density inhomogeneities affect the line source function and opacity (third and fourth panels of Fig.~\ref{fig:9}). These terms together with the continuum intensity contribute to Eq.~\ref{Eq:cfmagain}. This explains the splitting of the line formation depths seen in Fig.~\ref{fig:8}, and not seen in the 1D CFLD. 

\citet{2009A&A...506.1351C} showed that for some rays of the 3D simulation, the optical depth scale may vary by a large amount between two neighbor grid points (e.g., from $\tau=1$ to a few hundred). The intensity
is poorly estimated along those few rays. We tried to overcome this problem by interpolating the intensity, but without reliable results. 
Fortunately, the intensity only impacts the height of the 3D CFLD but does not affect the location of its maximum along the optical depth scale. Thus, the comparison with the 1D CFLD is not affected by the poor grid resolution of 3D simulations at large depths.

Figure~\ref{fig:9} shows that the temperature and density structures are obviously not similar along different single rays of the 3D simulation, and differ as well from those of a 1D model atmosphere. 
All rays of the 3D numerical box have to be taken into account, and the 3D $\rm CFLD_{DI}$ has to be computed, as we do in the following Section.

%                                                One column figure
%----------------------------------------------------------------- 
%----

\subsection{The CFLD for the 3D simulation: stellar disk}
\label{Sect: 3D-stellar-disk}

For the whole stellar disc, the derivation of the 3D $\rm CFLD_{DI}$ cannot be performed with Eq.~\ref{Eq:cfalbrow}, which describes a plane-parallel atmosphere. 

Following \citet{1996MNRAS.278..337A} and \citet{2015MNRAS.452.1612A}, we define the absolute depression in the line intensity as

\begin{equation}
Q_z(\tau) = I_{c,z}(\tau) - I_{l,z}(\tau)
\label{eq:5}
,\end{equation}
\
where the $z$ index means that the intensity is considered along the $z$ axis (pointing towards the observer). In the above relation, all quantities depend on wavelength $\lambda$ , and optical depth $\tau$, which is itself a function of the considered point in the atmosphere: $\tau = \tau(x,y,z)$;it is computed from $\tau(x,y,z)=\int_z^{z_s} \kappa(x,y,z') \, {\rm d} z'$ where $z_s$ is the $z$ coordinate at the surface.
\

The absolute line depression in the flux is then

\begin{equation}
U = F_c - F_l = \int_{\rm stellar~disk} Q(x,y,z_s) \, \frac{{\rm d} x {\rm d} y}{R^2}
\label{eq:6}
,\end{equation}
\
where
\begin{itemize}
\item {$F_c$ and $F_l$ are continuum and line fluxes respectively;}
\item {$Q(x,y,z_s)$ is computed at the surface;}
\item {$R$ is the stellar radius. }
\end{itemize}

{Since no flux is coming from the opposite side of the star, the integration in Eq.~\ref{eq:6} is performed over a hemisphere.}

{Since $Q$ obeys a transfer equation \citep[e.g., Eq. 7 of][]{1996MNRAS.278..337A}, its formal solution at the surface may be written:}

\begin{equation}
\label{Eq: Q-1}
Q(x,y,z_s) = \int_0^\infty S_Q(x,y,z) \, e^{-\tau} \, {\rm d} \tau
,\end{equation}
\
{with $z$ being an implicit function of $\tau$ as defined above, and with} 
\
\begin{equation}
\label{eq:9}
S_Q(x,y,z) = \frac{\kappa_l(x,y,z)}{\kappa_c(x,y,z) + \kappa_l(x,y,z)} (I_{c,z}(x,y,z) - S_{l,z}(x,y,z)).
\end{equation}

{
We note that Eqs.~\ref{eq:6} and ~\ref{Eq: Q-1} are equivalent to Eq.~11 of \citet{2015MNRAS.452.1612A}.
As compared to \citet{1996MNRAS.278..337A} formalism, there is no need for the variable $\mu$ in our treatment 
($\mu \equiv 1$ because all rays are parallel and in the direction of the observer).}

{Substituting $Q$ in the definition of $U$ then leads to}

\begin{equation}
U = \int_0^\infty \iint  \frac{{\rm d} x {\rm d } y}{R^2}  S_Q(x,y,z) \, e^{- \tau } \, {\rm d} \tau
.\end{equation}

{The contribution function to the line depression in the flux is therefore:}

\begin{equation}
\label{eq:11}
C_U (\tau) = \iint  \frac{{\rm d} x {\rm d } y}{R^2} S_Q(x,y,z(\tau)) \, e^{- \tau}.
\end{equation}

{Despite the fact that this function is applicable to a 3D model, it is nevertheless a function of the single variable $\tau$.}

\begin{figure}[h]
\centering
\includegraphics[width=9cm]{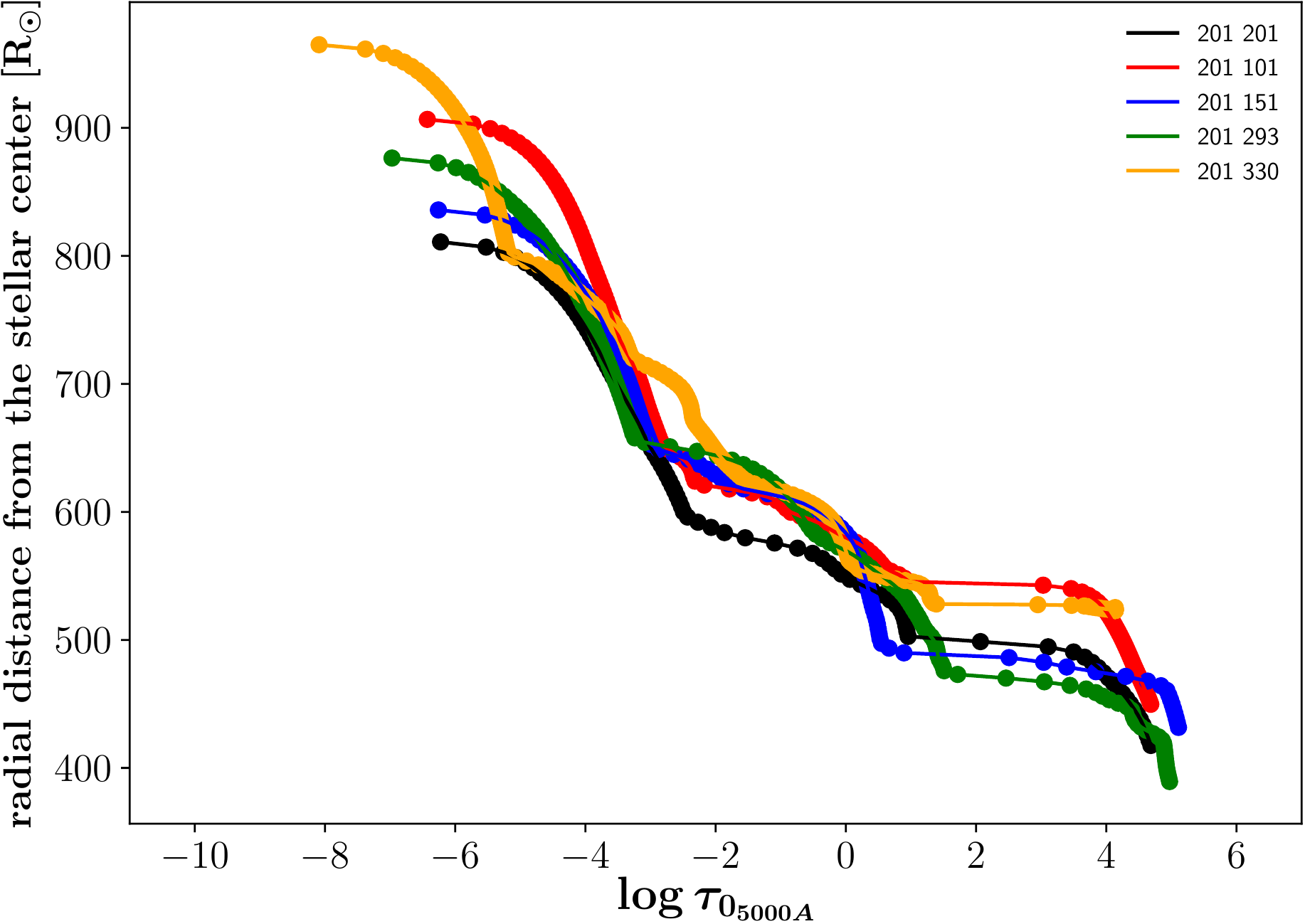}
\caption{Relation between the reference $\tau_0$ scale and radial distances (in $\rm R_{\odot}$) from points along five rays of the 3D simulation to the stellar center. Numbers on the labels correspond to $i$ and $j$ indices of selected rays.}
\label{fig:10}
\end{figure}

{The link between $\tau$ and the geometrical depth $z$ is however different for each of the considered parallel rays, described by the variables $x,y$. This is illustrated in Fig.~\ref{fig:10} and is caused by the dynamics that affects the opacity and structure of the atmosphere from ray to ray. For convenience, this diversity in the relationship between $\tau$ and $z$ depending on $(x,y)$ is illustrated by showing the spread in the run of $\tau$ with radial distance $r = (x^2+y^2+z^2)^{1/2}$ within the star for a few $(x,y)$ pairs (represented by the ray number $i$, $j$).}

{Equation~\ref{eq:11} may be discretized as follows:}

\begin{equation}
C_U (\tau) = \sum_{i,j=0}^{N} S_Q (x_i,y_j,z(\tau)) \, e^{-\tau } \frac{ \Delta x_i \Delta y_j}{R^2}
,\end{equation}
\
{and a reference $\tau_0$ scale may be defined (Eq.~\ref{Eq:tau}), so that}

\begin{equation}
C_U (\tau_0) = C_U (\tau) \, \frac{\kappa_l + \kappa_c}{\kappa_0} \, \tau_0 \, {\rm ln} 10
,\end{equation}
\
{and}

\begin{equation}
\begin{split}
C_U (\tau_0) = {\rm ln} 10 \sum_{i,j=0}^{N} \, \tau_0 \, \frac{\kappa_{l,i,j}}{\kappa_{0,i,j}} \, \bigg[I_{c,z} \Big(x_i,y_j,z(\tau_0)\Big) - \\
S_{l,z} \Big(x_i,y_j,z(\tau_0)\Big)\bigg] e^{-\tau } \frac{ \Delta x_i \Delta y_j}{R^2}
\end{split}
\label{Eq: finalCU}
,\end{equation}
\
{where $N$ is the number of grid points in the numerical box along the $z$ axis. Since the numerical grid is equidistant,  $\Delta x_i = \Delta y_j$ . In the above relation, all quantities depend moreover on $\lambda$.}

{We note that $C_U(\tau_0)$ fulfills the usual requirements for a contribution function, namely: $C_U(\tau_0) \neq 0$ only if $\kappa_l \neq 0$ and $S_l \neq I_c$, and $C_U(\tau_0)>0$ if $S_l < I_c$ (absorbing region).}
{In principle, the stellar radius needs to be computed to use Eq.~\ref{Eq: finalCU}; however, it acts as a normalization constant. Following the method explained in \citet{2011A&A...535A..22C}, the temperature and luminosity were averaged over spherical shells, and the stellar radius was computed using Eq.~5 from \citet{2011A&A...535A..22C}. It amounts to $582 \pm 5 \, R_{\odot}$ for the 3D simulation from Table~\ref{table:1}.}

%                                                One column figure
%----------------------------------------------------------------- 
%----

\begin{figure}
\centering
\includegraphics[width=9cm]{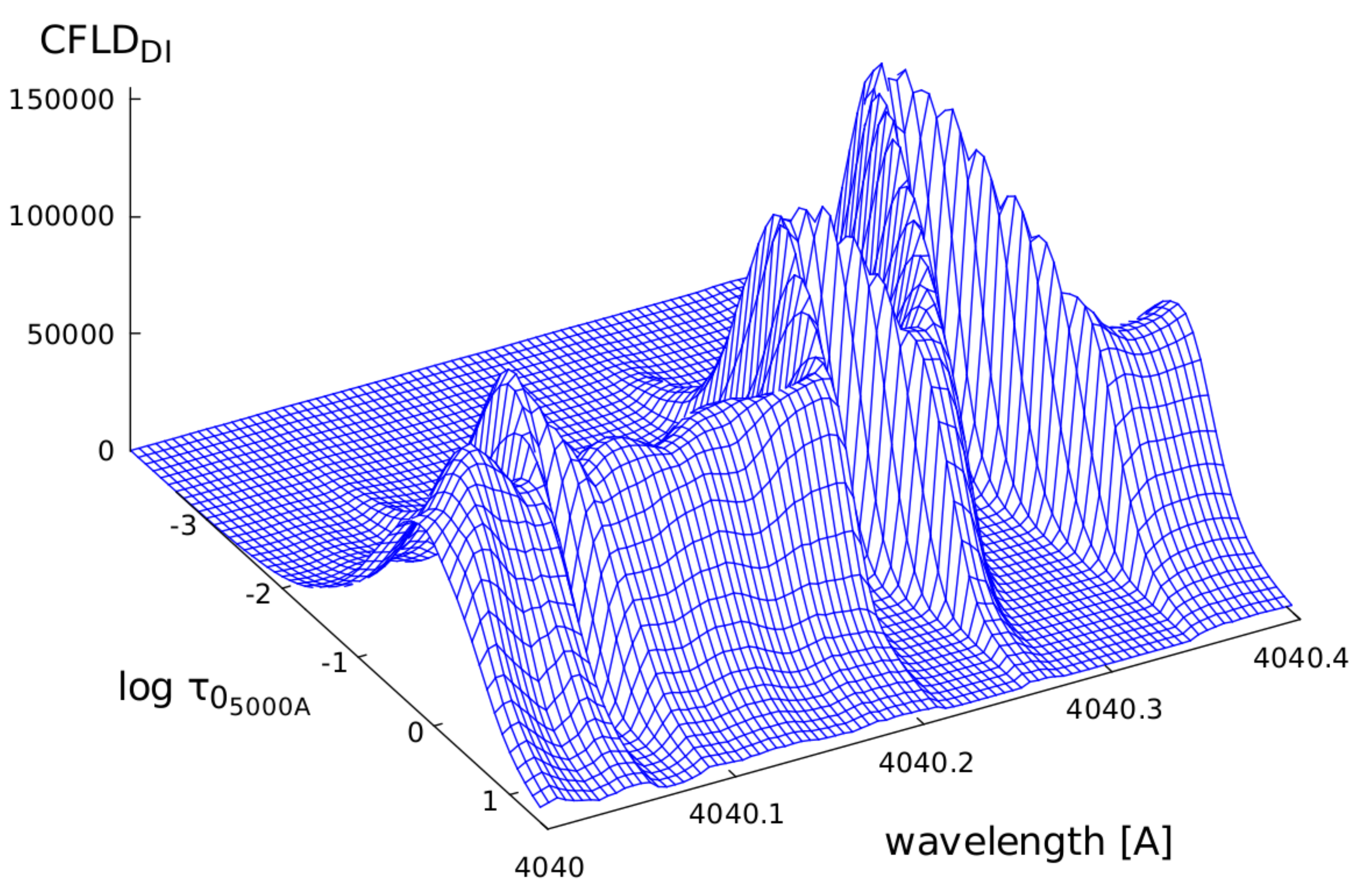}
\caption{The 3D $\rm CFLD_{DI}$ for the whole stellar disk, to be compared with top panel of Fig.~\ref{fig:1Dcfsurface}.  A spectral resolution of $R = 900000$ was adopted for plotting purposes.}
\label{fig:11}
\end{figure}

The 3D $\rm CFLD_{DI}$ was computed for the whole stellar disk using Eq.~\ref{Eq: finalCU}. W note that rays located in regions of the numerical box where the column density is so low that the integrated optical depth falls below the limit of $\tau = 2/3$ were not considered. The resulting 3D $\rm CFLD_{DI}$ is displayed in Fig.~\ref{fig:11} in the same spectral interval as the 3D $\rm CFLD_{SR}$ for the central ray of the 3D simulation (ray 1, see Fig.~\ref{fig:8}) and the 1D $\rm CFLD_{DI}$ (see top panel of Fig.~\ref{fig:1Dcfsurface}). The 3D $\rm CFLD_{DI}$ profiles are characterized by a single maximum in the $\log \tau_0 - \rm CFLD$ plane, similar to the 1D CFLD behavior. 

\begin{figure}
\centering
\includegraphics[width=8.5cm]{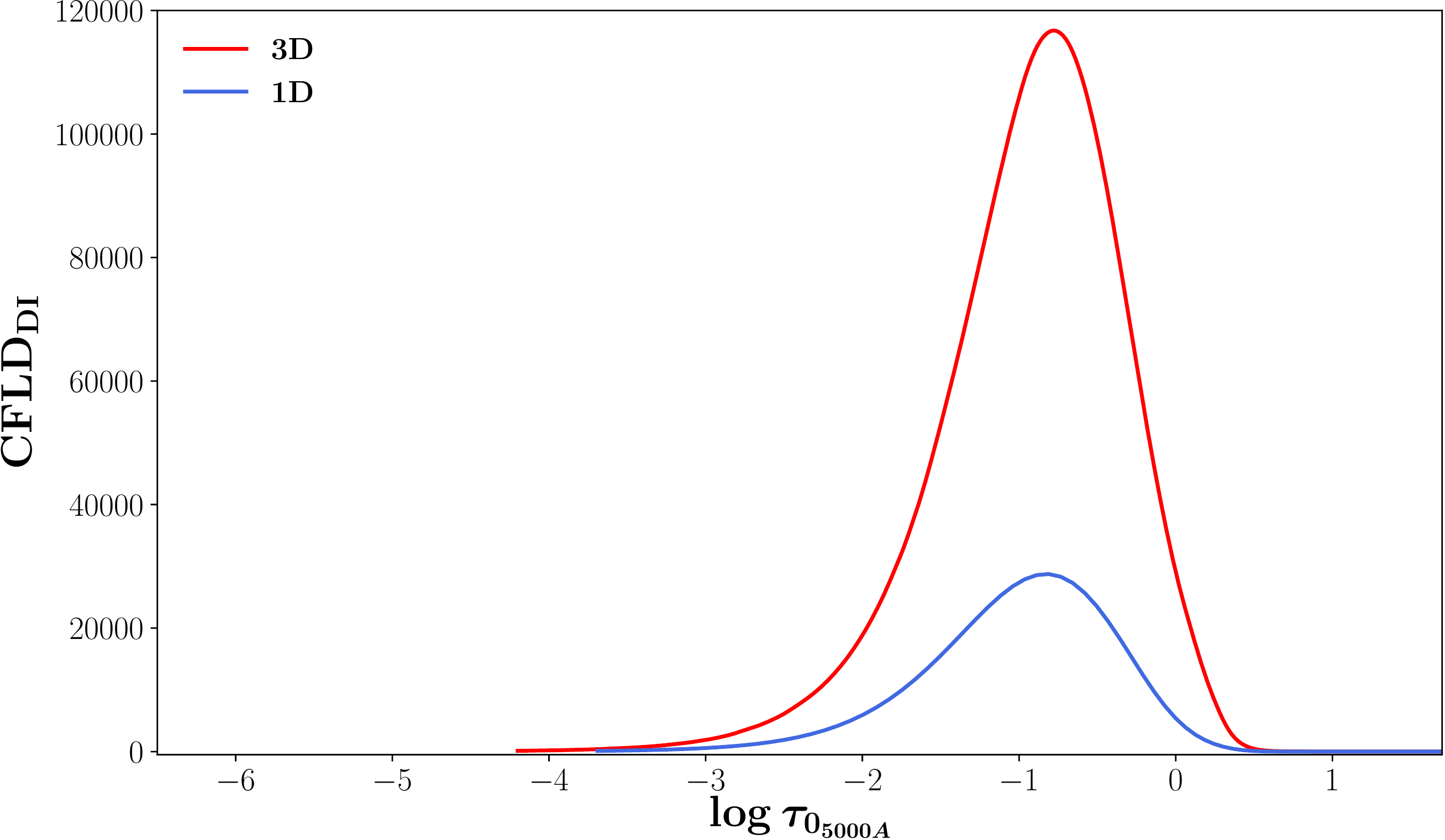} \\
\includegraphics[width=8.4cm]{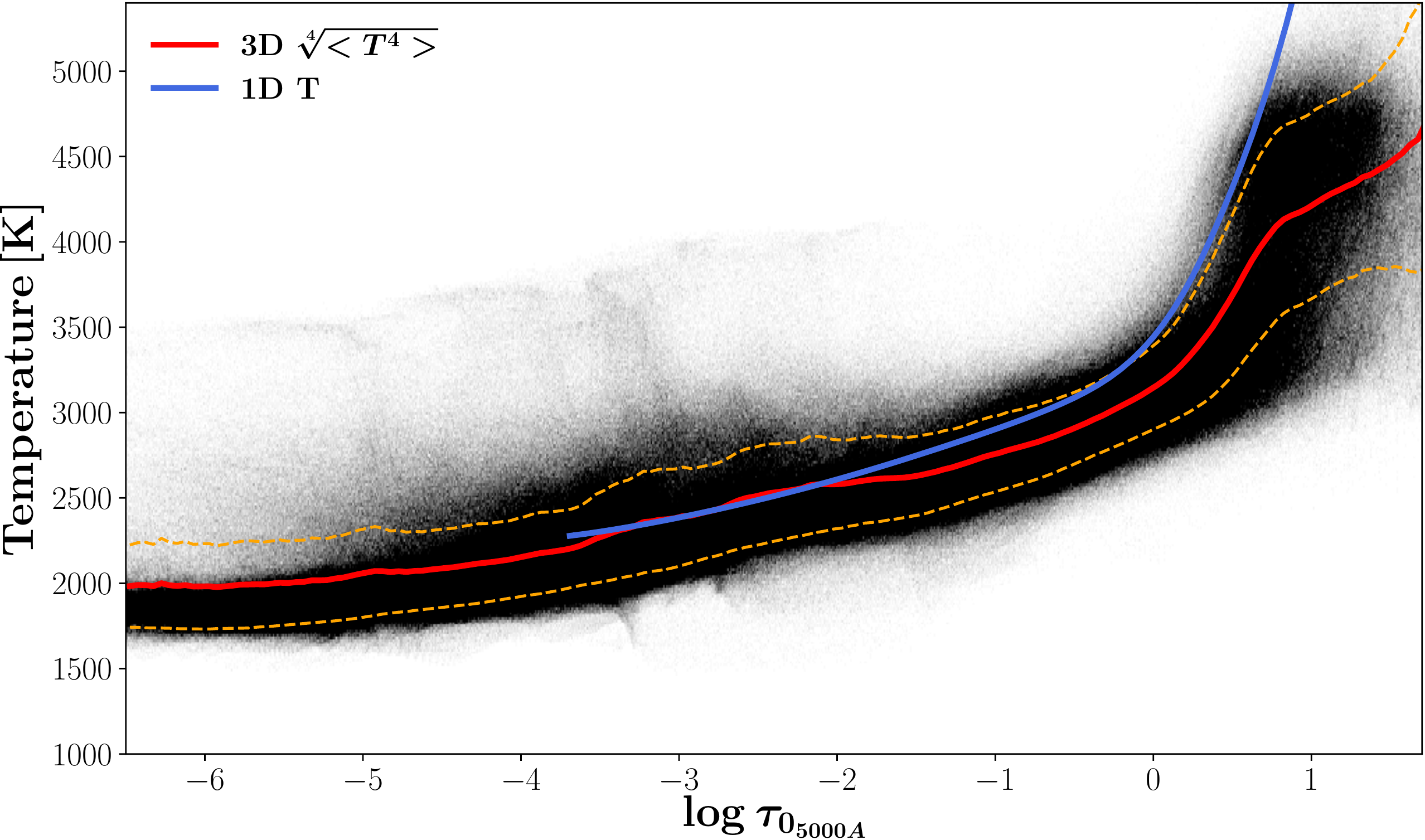}
\caption{\textit{Top panel:} 
The $\rm CFLD_{DI}$ for the whole stellar disk (red line) and for the 1D-RSG model of Table~\ref{table:1} (blue line) computed at $\lambda~=~{\rm 4040.07~\AA}$. \textit{Bottom panel:} 
The thermal structures for all rays contributing to the stellar disk. Darker areas correspond to more frequent temperature values. The red line is the average 3D thermal profile. The orange dashed lines show the $1 \sigma$ values around the average. The blue line is the 1D-RSG model thermal profile.}
\label{fig:12}
\end{figure}

The top panel of Fig.~\ref{fig:12} compares the 3D $\rm CFLD_{DI}$ at $\lambda =$ 4040.07 $\AA$ with the corresponding 1D $\rm CFLD_{DI}$ (computed using Eq.~\ref{Eq:cfalbrow}). 
A single-peaked 3D $\rm CFLD_{DI}$ is the result of a monotonic behavior of the average temperature profile of the 3D numerical box (see bottom panel of Fig.~\ref{fig:12}). Since many rays of the 3D simulation are taken into account, the atmospheric inhomogeneities largely average out, so that the global disk produces a 3D CFLD resembling a 1D one.

\subsection{Tomographic masks}
\label{Sect: tomo-masks}

{This Section aims at validating the fact that tomographic masks computed from a 1D \textit{static} model atmosphere do not lose their discriminating properties in a dynamical atmosphere. In other words, we need to check that a set of lines that originate from the same given depth in a static atmosphere (thus defining a \textit{mask}) continue to form in the {same} layers in a dynamical atmosphere. Of course, these layers will not necessarily be at the same depth in the dynamical atmosphere, and multiple layers may contribute to a given line (because the run of the temperature is no longer monotonic). But the key property to be satisfied is that all lines defining a given mask from a 1D static atmosphere behave {similarly} in a dynamical atmosphere. Moreover, we need to define the masks from a static atmosphere to be able to subsequently reveal complex matter motions in a dynamical atmosphere (showing up as multiple peaks in the 3D CFLD and corresponding CCFs).}

\begin{figure}
%\centering
\includegraphics[width=5.6cm]{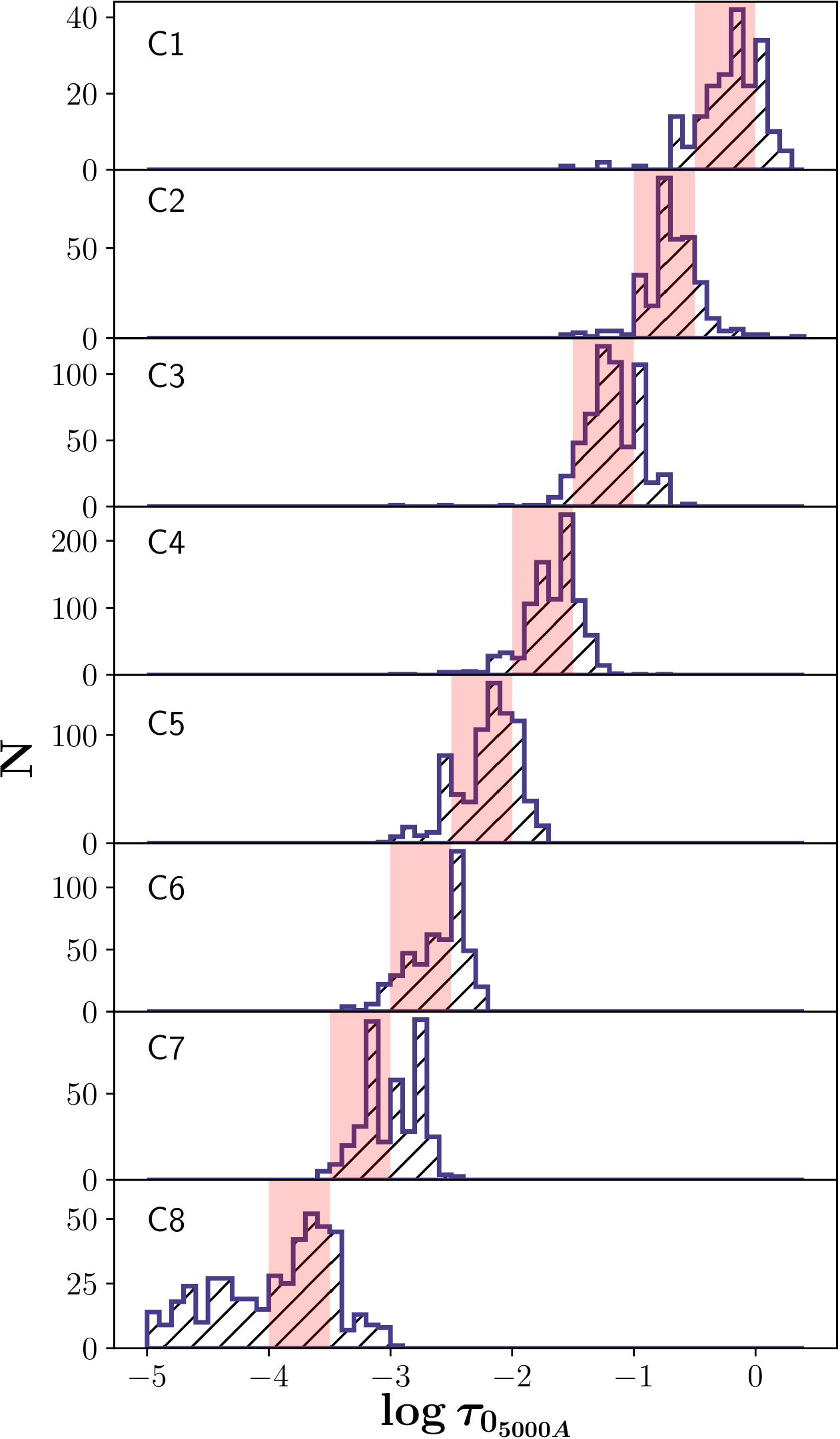} 
\includegraphics[width=3.2cm,height=9.57cm]{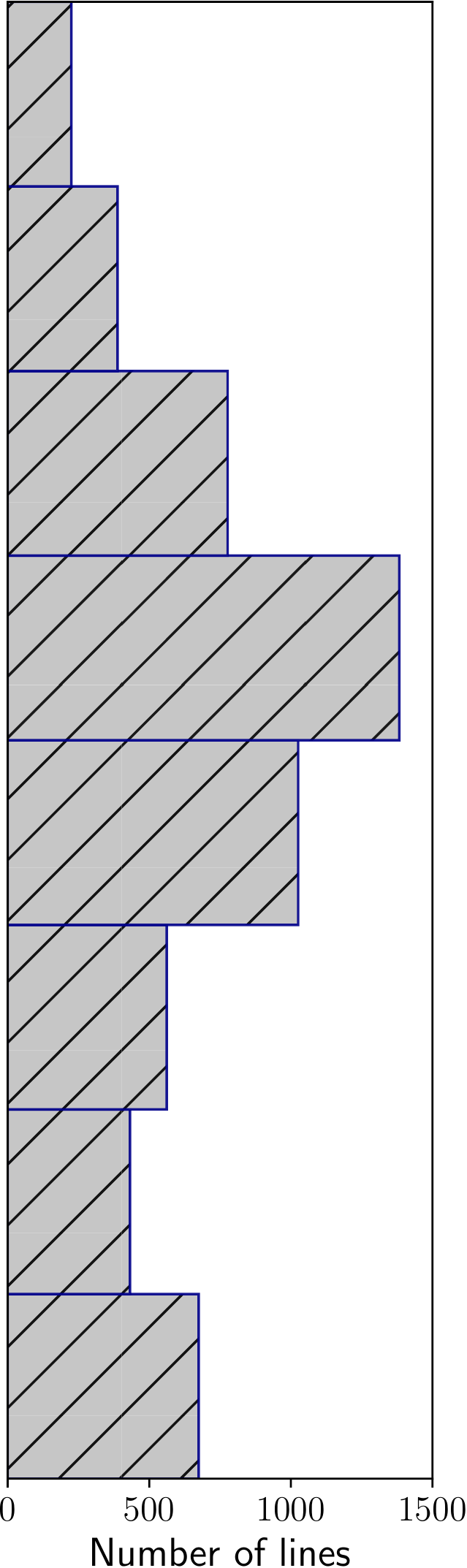}
\caption{Properties of masks C1-C8 computed from the 1D-RSG model (Table~\ref{table:1}). \textit{Left panel:} The distribution of formation depths of lines contributing to tomographic masks C1-C8 for the whole stellar disk (blue histograms). Red bands represent the optical depth ranges used for selecting lines in a given mask from the 1D CFLD. \textit{Right panel:} The distribution of the number of lines in spectral masks C1-C8.}
\label{fig:13}
\end{figure}

{In order to compare the line formation depths between the 1D model atmosphere and the 3D simulation, the set of eight tomographic masks constructed in Sect.~\ref{Sect: mask-construction} from a 1D RSG model atmosphere (Table~\ref{table:1}) were used. }
The distribution of the number of lines in masks is shown in the right panel of Fig.~\ref{fig:13}. The optical depths corresponding to maxima of the 3D $\rm CFLD_{DI}$ in a $\log \tau_0 - \rm{CFLD}$ plane for all lines contributing to a given mask were extracted and displayed as histograms on the left panel of Fig.~\ref{fig:13}. Red-colored areas indicate regions probed by masks according to the 1D model atmosphere.   

The histograms in Fig.~\ref{fig:13} move along the optical depth axis from the deepest layers (C1) to the outermost (C8) in the same manner as 1D line formation depths. The important conclusion at this point is that the line formation depths in the 3D model stay within the range expected from the 1D model and do not get scattered all over the $\tau_0$ scale for any given mask. Thus, masks constructed from a 1D model keep their power to probe different atmospheric depths in a dynamic atmosphere of the kind we used here. Mask C8 is characterized by a shift of line formation depths towards lower optical depths (outer layers). It is related with the fact that the 3D model atmosphere is more extended than the 1D model atmosphere.    
This is due to the turbulent pressure naturally included in 3D simulations that lowers the effective gravity and consequently stretches the atmosphere \citep{1992iesh.conf...86G,2008A&A...486..951G,2011A&A...535A..22C}. 

%                                                One column figure
%----------------------------------------------------------------- 
%----

%                                                One column figure
%----------------------------------------------------------------- 
%----

%                                                One column figure
%----------------------------------------------------------------- 
%----

\subsection{Can tomography reliably recover the $V_z$ distribution in a 3D atmosphere?}
\label{Sect: V-recovery}

\subsubsection{Single ray}
\label{Sect:v-recovery-singleray}

Previous sections showed that the line formation process in a 3D atmosphere is more complex than in a 1D model atmosphere (as seen in Figs.~\ref{fig:7} and \ref{fig:9}) leading to a small but real spread of line-forming regions in the atmosphere. 
This means that a given line will contribute to the CCF of its corresponding mask with different velocities. 
The resulting CCF will have multiple peaks which are the signature of the presence of a velocity field in the atmosphere. 
This Section aims at confirming that the peaks in the CCF are representative of the velocities. For this purpose, the 3D simulation is of invaluable help since the velocity value at each grid point of the numerical box is obviously known.

\begin{figure*}%[h]
\includegraphics[width=5.7cm]{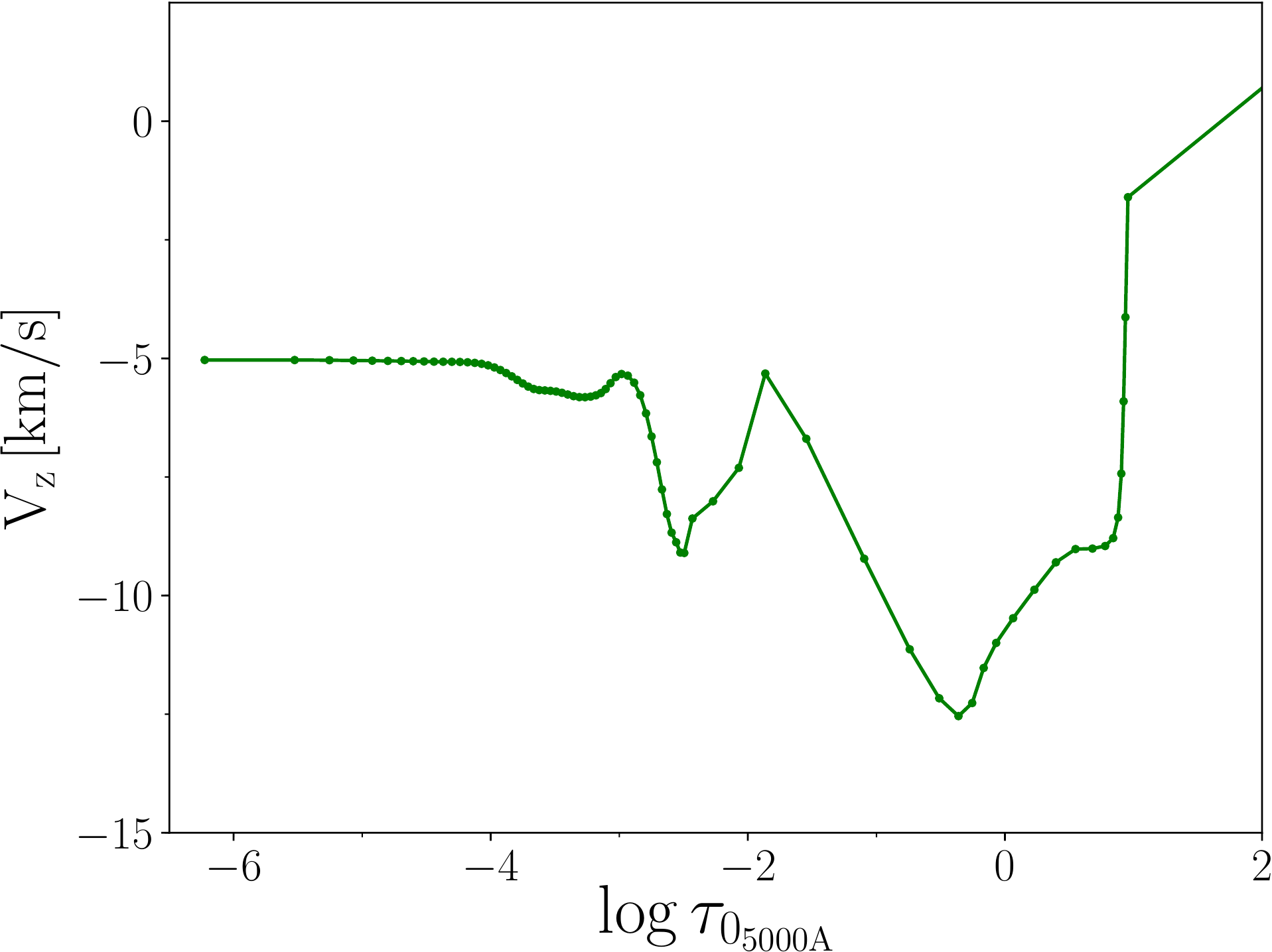} 
\includegraphics[width=6.cm]{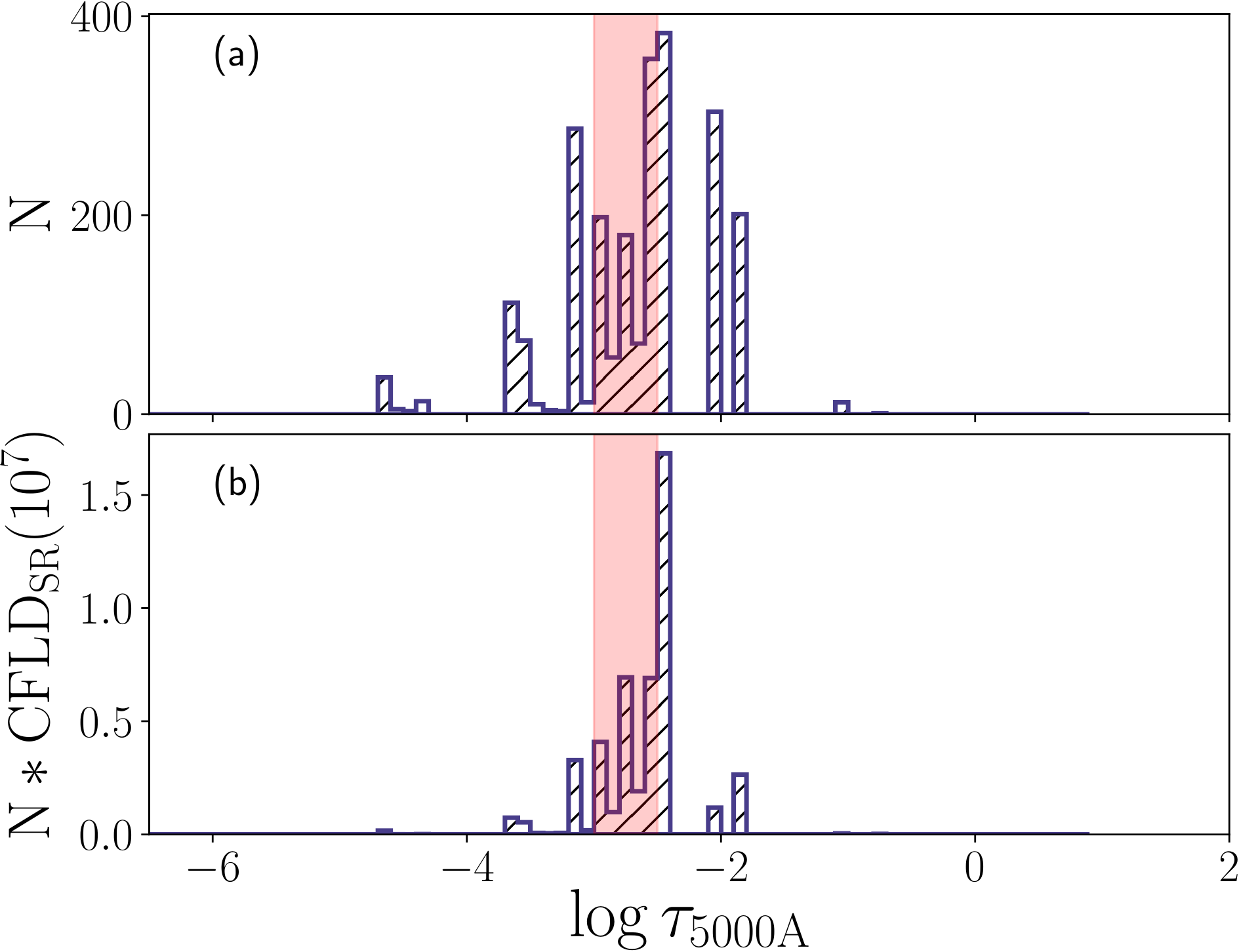} 
\includegraphics[width=6.1cm]{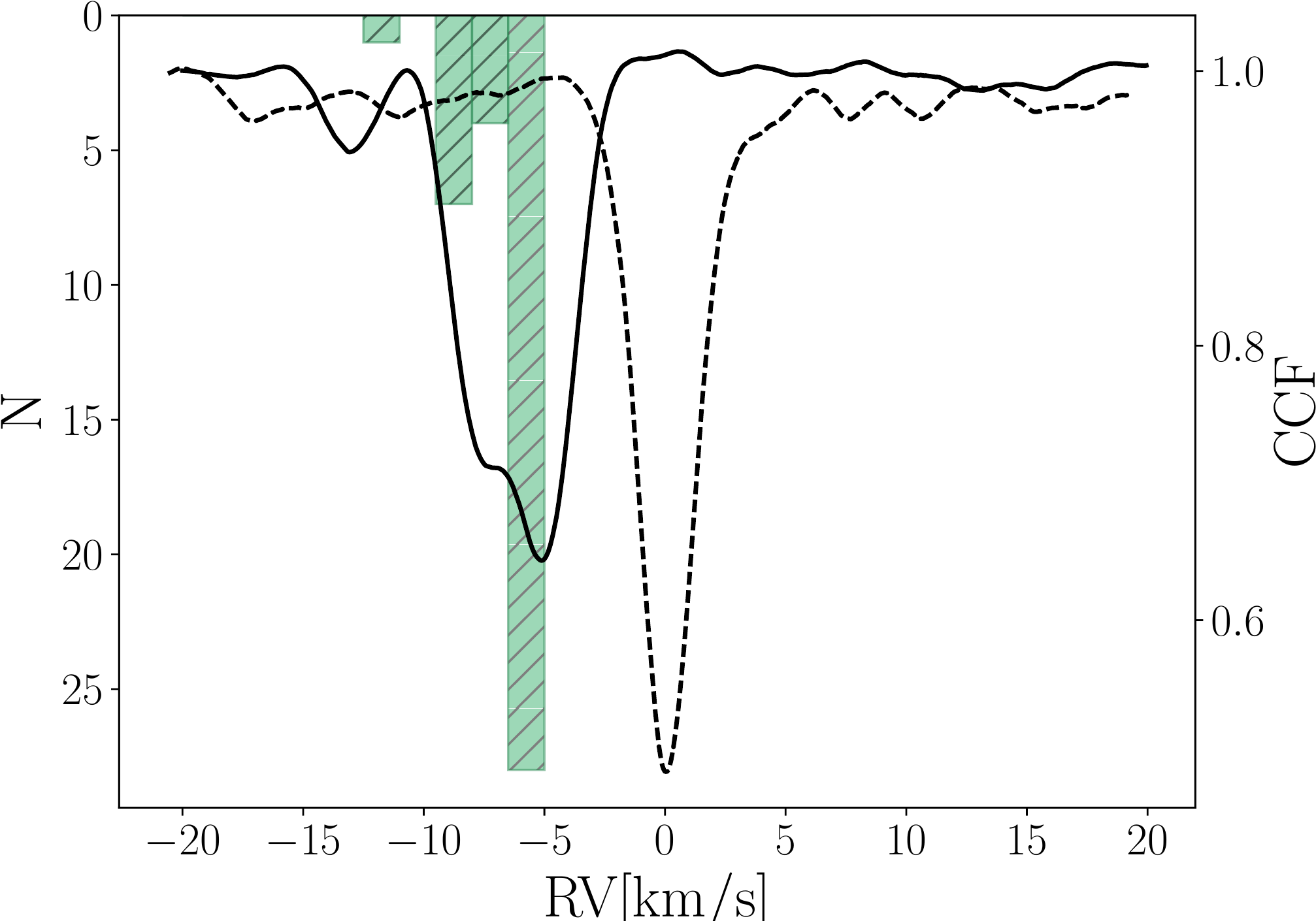} 
\caption{\textit{Left panel:} $V_z$ as a function of the reference optical depth for the ray 1 of the 3D simulation. \textit{Middle panel:} The distribution of formation depths of lines contributing to the mask C6 for the ray 1 weighted (b) and not weighted (a) by the CFLD. \textit{Right panel:} The CCF obtained by cross-correlation of the synthetic spectrum [\textit{with} (black solid line) and \textit{without} (black dashed line) including the velocity field in the 3D simulation] for the ray 1 of the 3D snapshot with the mask C6. Green bars show the distribution of velocities corresponding to formation depths of lines contributing to the mask C6.}\label{fig:14}
\end{figure*}

For this analysis, a single ray of the 3D numerical box (ray 1) was chosen, and a synthetic spectrum was computed with a spectral resolution $R = \lambda / \Delta \lambda =$ 200000, including and not including the velocity field in the 3D simulation. Then, it was cross-correlated with the tomographic mask C6 obtained in Sect.~\ref{Sect: mask-construction}. The right panel of Fig.~\ref{fig:14} displays the resulting cross-correlation profile with and without the velocity field. The CCF has three components, at -6, -9 and -12 km/s. The CCF produced with all velocities at zero is useful to measure the level of correlation noise (caused by the forest of neighboring lines).

As a next step, the optical depths corresponding to all maxima of the 3D $\rm CFLD_{SR}$ for all {wavelengths} contributing to the mask C6 were retrieved. They are displayed as histograms on the middle panel of Fig.~\ref{fig:14} weighted (b) and not weighted (a) by the respective $\rm CFLD_{SR}$ value. Weighting of the histogram by the CFLD shows which peaks on the CFLD significantly contribute to the resulting intensity used for the CCF computation.

Then, the $V_z$ velocities along a given ray (left panel of Fig.~\ref{fig:14}) attributed to the optical depth ranges from the middle panel of Fig.~\ref{fig:14} were extracted. They are displayed as histograms (green color) in the right panel of Fig.~\ref{fig:14} with the binning step equal to 1.5 km/s to match the resolution of the CCF. The histograms have peaks at -6, -9 and -12 km/s which are present on the CCF. This shows that our method correctly recovers the velocity value of all components on the CCF.

\subsubsection{Integrated flux}
\label{Sect: 3D-V-recovery}

\begin{figure}%[h]
\centering
\includegraphics[width=8.5cm]{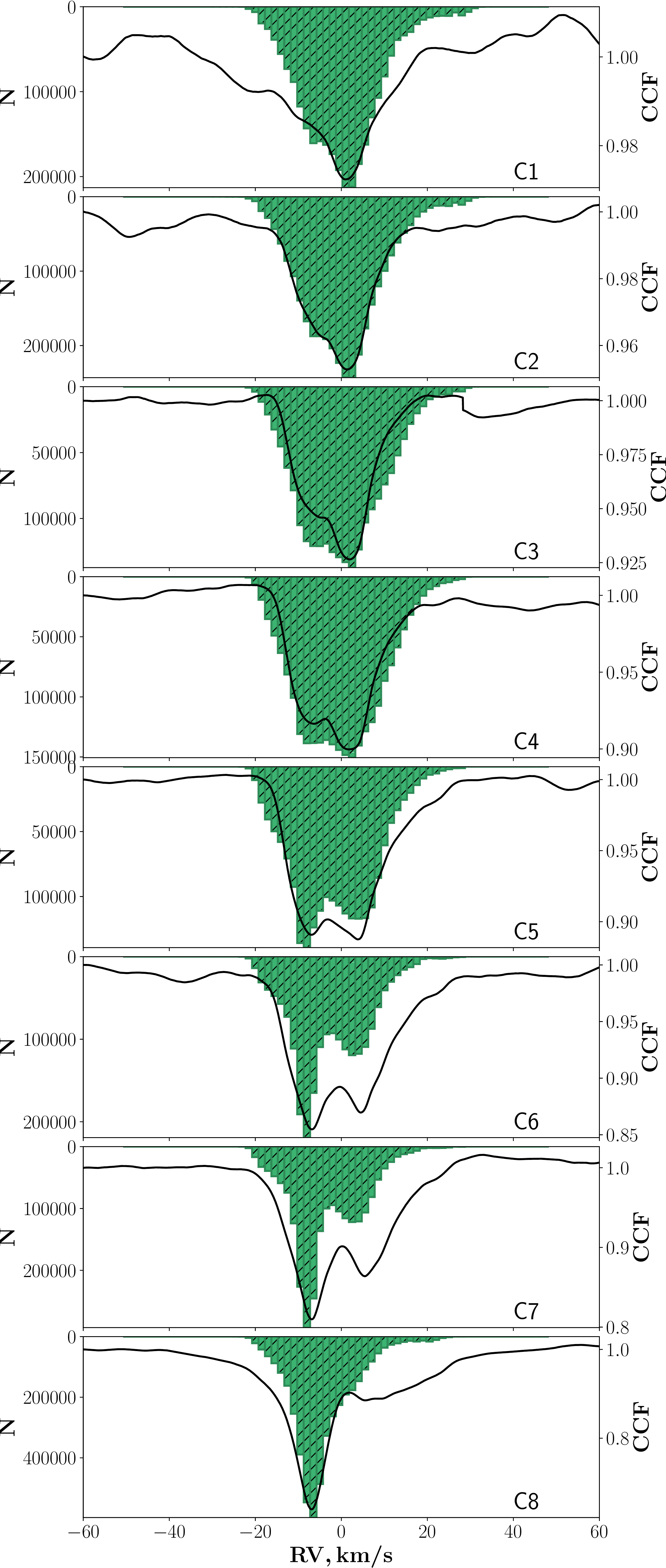} 
\caption{Sequence of CCF profiles (black solid line) obtained by cross-correlation of a synthetic spectrum (with included velocity field) for the whole stellar disk with tomographic masks C1-C8. Green bars show the distribution of velocities corresponding to all formation depths of lines contributing to masks C1-C8.}
\label{fig:15}
\end{figure}

The same analysis was performed for the whole 3D simulation. Figure~\ref{fig:15} shows the CCFs and the velocity histograms computed in the same way as those for a single ray described above. Here, the CCFs in the innermost masks C1-C2 have a main peak at 2 km/s. In mask C3, a blue-shifted component is clearly visible at -8 km/s. Moving to outermost mask C8, the strength of the blue-shifted peak increases, while that of the red-shifted peak decreases and shifts to 5 km/s. The evolution of peaks on the CCFs from red-shifted to blue-shifted while probing more and more external layers does not follow the Schwarzschild scenario, which describes the global envelope pulsation in Mira variables. In supergiant stars instead, the tomographic method probes local motions involving specific rays.

\begin{figure}%[h]
\centering
\includegraphics[width=8cm]{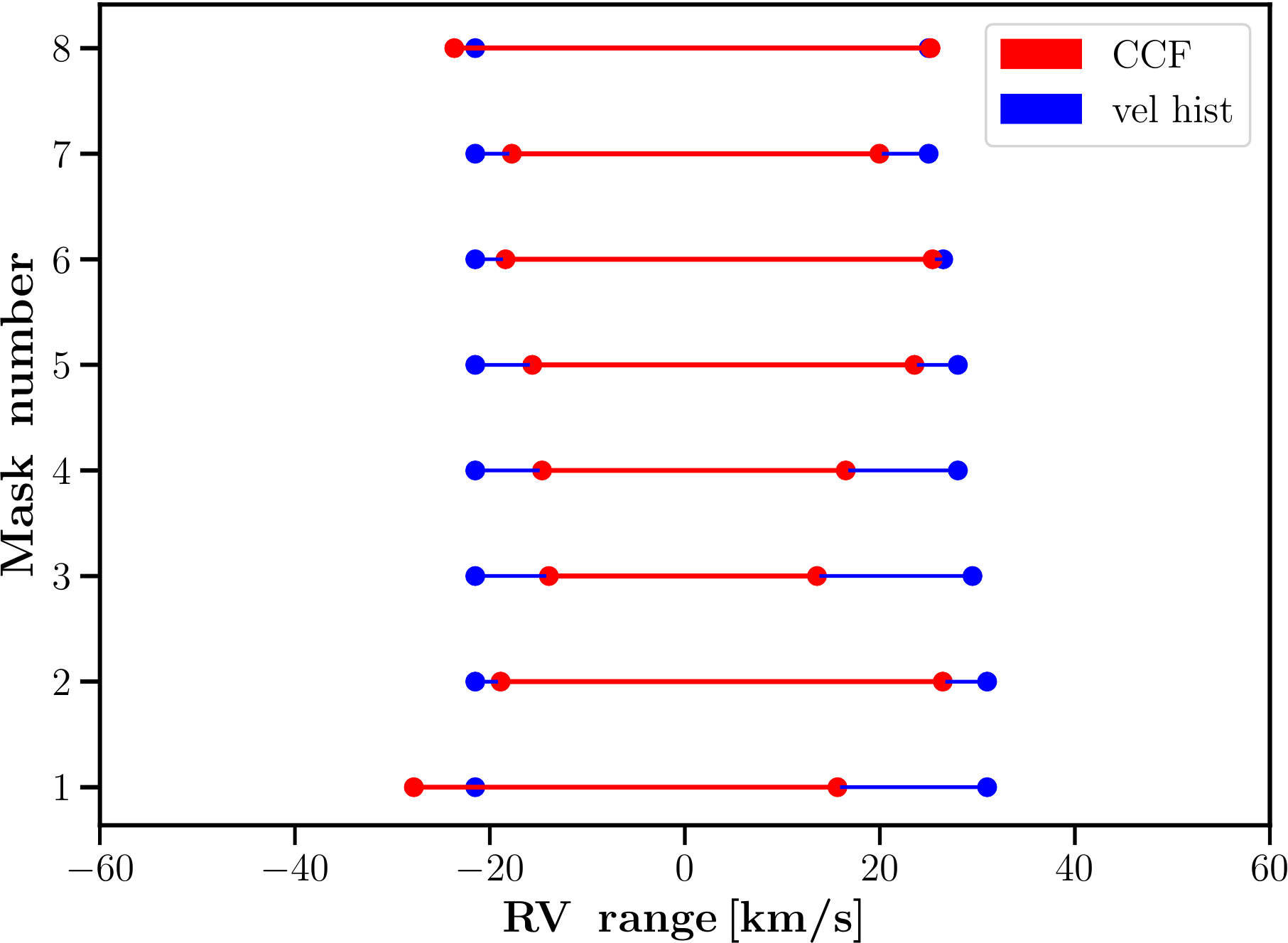} \\
\includegraphics[width=8cm]{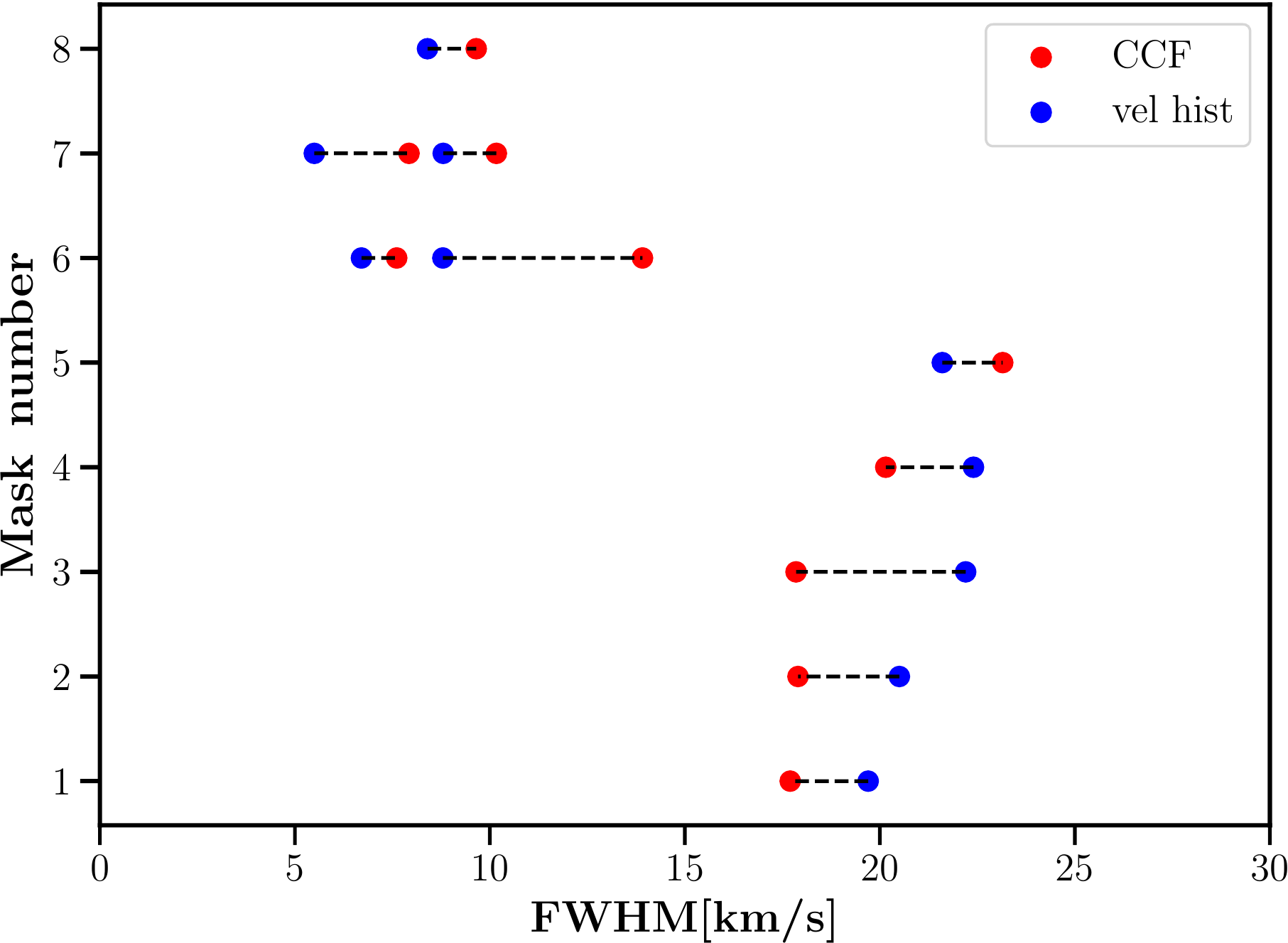} 
\caption{\textit{Top panel:} the velocity range covered by CCFs (red color) and velocity histograms (blue color) from Fig.~\ref{fig:15} in each spectral mask. \textit{Bottom panel:} the FWHM of CCF profiles (red color) and velocity distributions (blue color) for all masks. In masks C6 and C7 the FWHM were measured separately for each of the two peaks. The dashed black line connects the FWHM values of corresponding peaks.  }
\label{fig:16}
\end{figure}

In order to quantify the agreement between the velocity distributions and the CCFs, we first measured the velocity range they cover in each mask. In order to determine that velocity range for CCFs, we assume that the level of the correlation noise does not exceed 10\% of the total contrast of a given CCF. The velocity spread is displayed in the top panel of Fig.~\ref{fig:16} and is roughly similar for the CCFs and the histograms. Therefore, the CCFs correctly reproduce the range of velocities of the atmosphere. As a next step, the FWHM of velocity distributions and CCF profiles were computed. They are shown on the bottom panel of Fig.~\ref{fig:16}. Despite a slight difference in the FWHM of the CCFs and the histograms (especially for masks C3, C4, C6, and C7), we may nevertheless consider that the tomographic method is able to correctly recover the $V_z$ distribution.

\begin{figure}%[h]
\centering
\includegraphics[width=8.5cm]{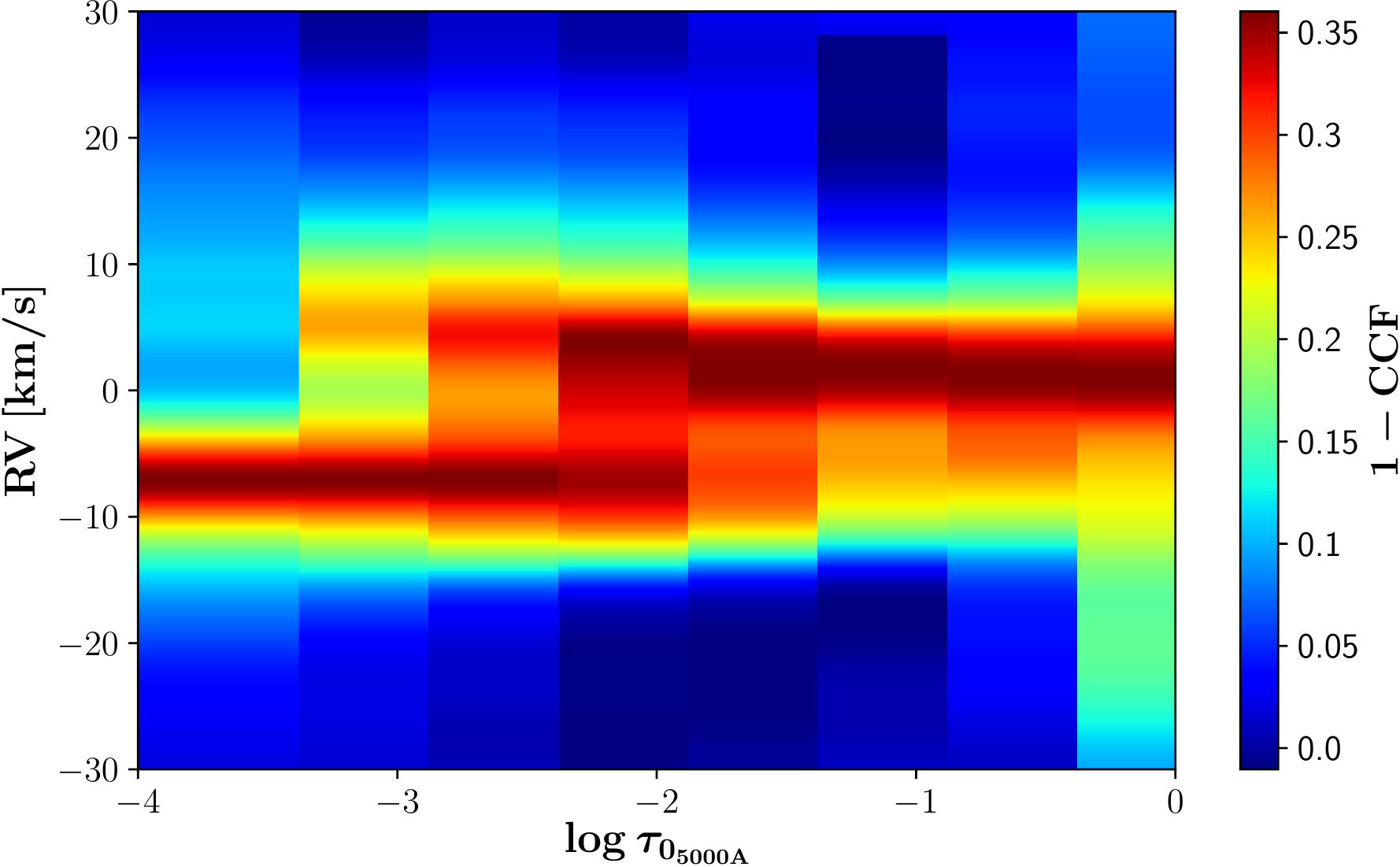} \\
\includegraphics[width=8.5cm]{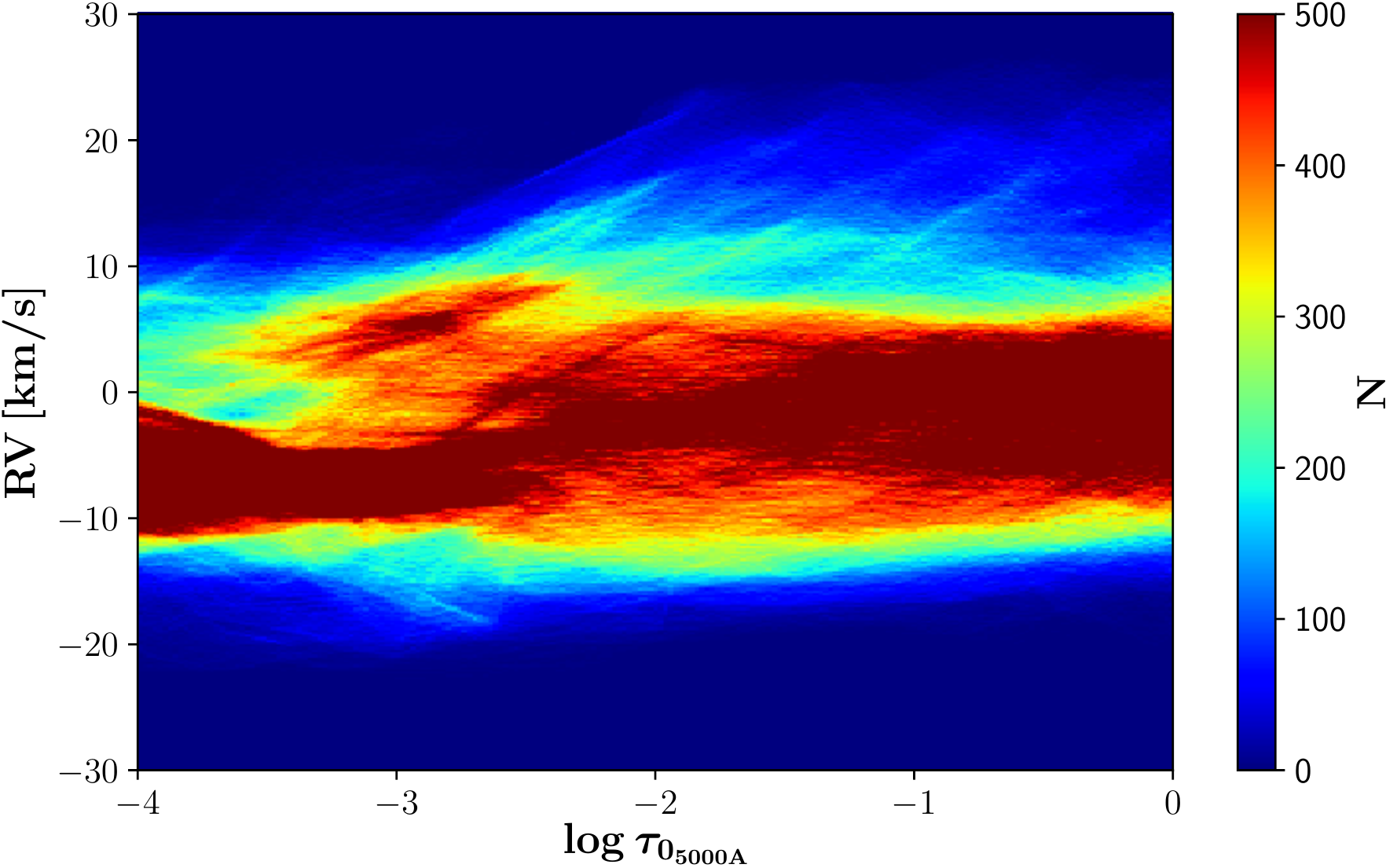}
\caption{\textit{Top panel:} The CCFs computed on a snapshot spectrum from the 3D simulation as a function of the reference optical depth scale and the velocity. The color code shows the minimum (red color) and maximum (blue color) values of each CCF. \textit{Bottom panel:} The distribution of $V_z$ velocities of the 3D simulation as a function of the reference optical depth. The color code shows areas with high (red) or low (blue) density of points.}
\label{fig:17}
\end{figure}

The top panel of Fig.~\ref{fig:17} displays the CCFs for each spectral mask as a function of the $\log \tau_0$ range and the velocity. The comparison with the distribution of velocities of the 3D simulation (the bottom panel of Fig.~\ref{fig:17}) shows that CCFs recover well the distribution of velocities.

%                                                One column figure
%----------------------------------------------------------------- 
%----

\section{Conclusions}
\label{Sect: Conclusion}

The present paper applies tomography to red supergiant atmospheres, aiming at recovering the projected velocity field at different optical depths in the stellar atmosphere. Our implementation of tomography includes the computation of the contribution function to the line depression to correctly assess the formation depth of spectral lines. 
{The method was compared to a simpler procedure proposed by \citet{2001A&A...379..288A}. We reproduce these earlier results for the Mira variable V Tau, thus validating all the conclusions from this earlier study.}

The tomography was applied to 3D RHD simulations in order to check whether the $V_z$ velocity fields in their atmospheres could be recovered by the method. For this purpose, the {CFLD} was calculated for individual rays of the 3D simulation and for the whole stellar disk. It showed that in 3D simulations the spectral lines do not form in a restricted range of reference optical depths as in 1D model atmospheres. The line formation is spread over different optical depths due to convection present in 3D simulations.

By comparing the CCF computed from the synthetic spectra originating from 3D simulations with the velocity distribution (available
from the very same simulations), it was shown that the CCFs indeed nicely trace the projected velocity fields in the atmosphere.

The application of the tomographic method to observed spectra of actual red supergiant stars is deferred to a forthcoming paper.

%                                                One column figure
%----------------------------------------------------------------- 
%----

%                                                One column figure
%----------------------------------------------------------------- 
%----

%                                                One column figure
%----------------------------------------------------------------- 
%----

\begin{acknowledgements}
K.K. acknowledges the support of a FRIA (FNRS) fellowship. S.V.E. is supported by a grant from the Fondation ULB. This work was granted access to the HPC resources of "Observatoire de la C\^ote d'Azur - M\'esocentre SIGAMM".
We wish to thank the referee Amish Amarsi for helpful comments which contributed to an improvement of the paper.
\end{acknowledgements}

\bibliographystyle{aa} % style aa.bst

\bibliography{bibliography.bib} % your references Yourfile.bib

\end{document}